\documentclass[a4paper,preprint]{revtex4-1}
\usepackage{amsmath,amssymb,amsfonts,latexsym,graphicx}
\usepackage[dvips]{color}
\usepackage{indentfirst}

\newcommand{\vect}[1]{\textbf{\textit{#1}}}

\begin{document}
\title{Path Integral Molecular Dynamics within the Grand Canonical-like Adaptive Resolution Technique: Simulation of Liquid Water}
\author{Animesh Agarwal}
\email{animesh@zedat.fu-berlin.de}
\affiliation{Institute for Mathematics, Freie Universit\"at Berlin, Germany}
\author{Luigi Delle Site}
\email{dellesite@fu-berlin.de}
\affiliation{Institute for Mathematics, Freie Universit\"at Berlin, Germany}

\begin{abstract}
Quantum effects due to the spatial delocalization of light atoms are treated in molecular simulation via the path integral technique. Among several methods, Path Integral (PI) Molecular Dynamics (MD) is nowadays a powerful tool to investigate properties induced by spatial delocalization of atoms; however computationally this technique is very demanding. The abovementioned limitation implies the restriction of PIMD applications to relatively small systems and short time scales. One possible solution to overcome size and time limitation is to introduce PIMD algorithms into the Adaptive Resolution Simulation Scheme (AdResS). AdResS requires a relatively small region treated at path integral level and embeds it into a large molecular reservoir consisting of generic spherical coarse grained molecules.
It was previously shown that the realization of the idea above, at a simple level, produced reasonable results for toy systems or simple/test systems like liquid parahydrogen.
Encouraged by previous results, in this paper we show the simulation of liquid water at room conditions  where AdResS, in its latest and more accurate Grand-Canonical-like version (GC-AdResS),  is merged with two of the most relevant PIMD techniques available in literature. The comparison of our results with those reported in literature and/or with those obtained from full PIMD simulations shows a highly satisfactory agreement. 
\end{abstract}

\maketitle

\section*{Introduction}
The structure and dynamics of liquids consisting of molecules that contain light atoms (e.g. hydrogen) can be influenced by the quantum effects due to the delocalization of atoms in space. In simulation such systems are treated by modeling the atoms of the molecules via the path integral formalism of Feynman \cite{feynbook1,feynbook2, tuckbook}.
In particular liquid water is a typical subject of interest given its role in many fields \cite{ball}. As explained more in detail in next sections, the computational effort is massive because the number of interatomic interactions becomes much larger compared to the classical case. As a consequence the size of the system and the simulation time affordable with standard computer resources is rather limited.
For liquid water at room condition a system of 500 molecules for a simulation time of 1-2 $ns$ is usually considered already expensive. The limited size and simulation time may imply that particle number density fluctuations are arbitrarily suppressed and some systems cannot be treated if not at high computational prize (e.g. solvation of a large molecule in water). An optimal complementary technique would consist of a Grand Canonical-like scheme where (local) properties can be calculated by employing a computationally affordable path integral simulation of a small open region which, in statistical and thermodynamic equilibrium, exchanges particles and energy with a reservoir acting at small computational cost. One possible implementation of a Grand Canonical-like Molecular dynamics technique is the Adaptive Resolution Simulation scheme (AdResS) \cite{jcp1,annrev,jcpsimon} in its most accurate version of GC-AdResS \cite{prlgio,jctchan,prx,jcpanim,njpyn}.
For the simplest version of AdResS it was shown that for a toy system (liquid of tetrahedral molecules) the embedding of a PIMD technique into the scheme produced rather encouraging results \cite{prlado}; such results were confirmed and empowered by the application to simple/test systems like liquid parahydrogen at low temperature \cite{para1,para2}. In the meanwhile the increased accuracy and more solid conceptual framework of the adaptive scheme (GC-AdResS) allows for the study of more complex systems and the calculation of a larger number of properties than before \cite{jctchan,prx,jcpanim,njpyn}. In this perspective, this paper reports the technical implementation of two different approaches to PIMD, Refs.\cite{tuck1, voth, marx} and Refs.\cite{man1, man3}, into our GC-AdResS. We show its application to liquid water and report results about static and dynamic properties. The comparison with reference data is highly satisfactory and suggest that GC-AdResS, as a complementary method, may play an important role in future applications of PIMD (today not feasible with full PIMD simulations). One can think, for example, of solvation of a large molecules (e.g. fullerene in water) and look at possible quantum effects in the structure of the solvation shell. Moreover, GC-AdResS may be employed as a tool of analysis and study how the quantum effects change as a function of the size of the region treated at PI level. This would represent a novel type of analysis because it  unequivocally defines the essential molecular degrees of freedom required for a given property \cite{fullerene} and thus it allows to quantify how localized are (possible) quantum effects (for the properties considered). The paper is organized as follows: next section is dedicated to a summary of the relevant technical and conceptual characteristics of GC-AdResS, it follows the section dedicated to the description of the basic characteristics of the two PIMD methods employed in this study. Next the implementation of PIMD in GC-AdResS, for each of the two specific techniques used, is reported. It follows the section of results divided into the subsection of (i) static and (ii) dynamic properties. In (i) we report particle number density profiles, probability distributions and radial distribution functions of the GC-AdResS simulation compared with results from full PIMD simulations. In (ii) we report the calculation of equilibrium time correlation functions compared, also in this case, with data obtained from full PIMD simulations.
Finally the section of discussion and conclusion is presented. The appendix instead reports all technical data of the simulations so that the results can be reproduced/checked by other groups.  
\section*{ GC-AdResS}
\begin{figure}
  \centering
  \includegraphics[width=0.75\textwidth]{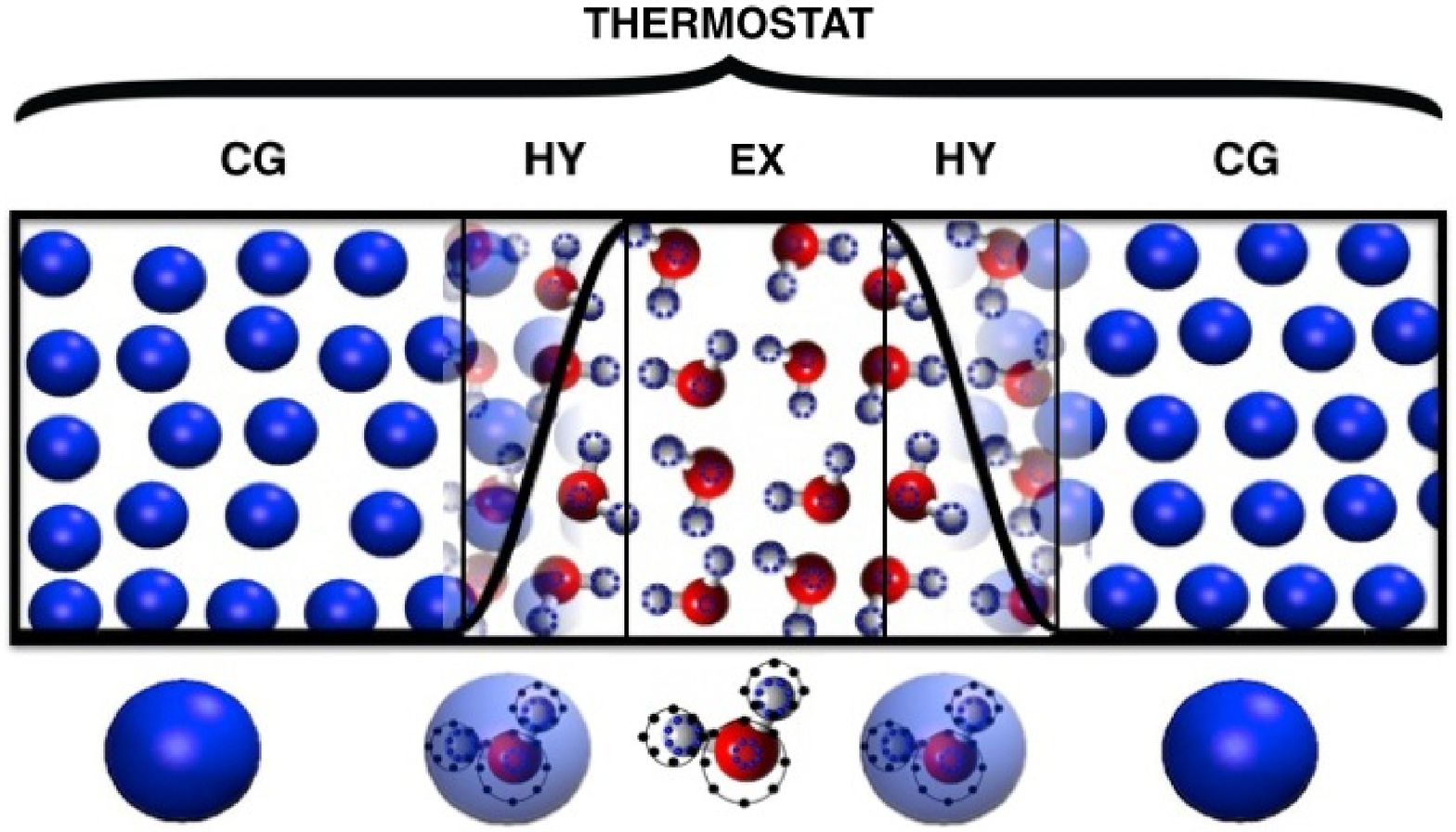}
  \includegraphics[width=0.75\textwidth]{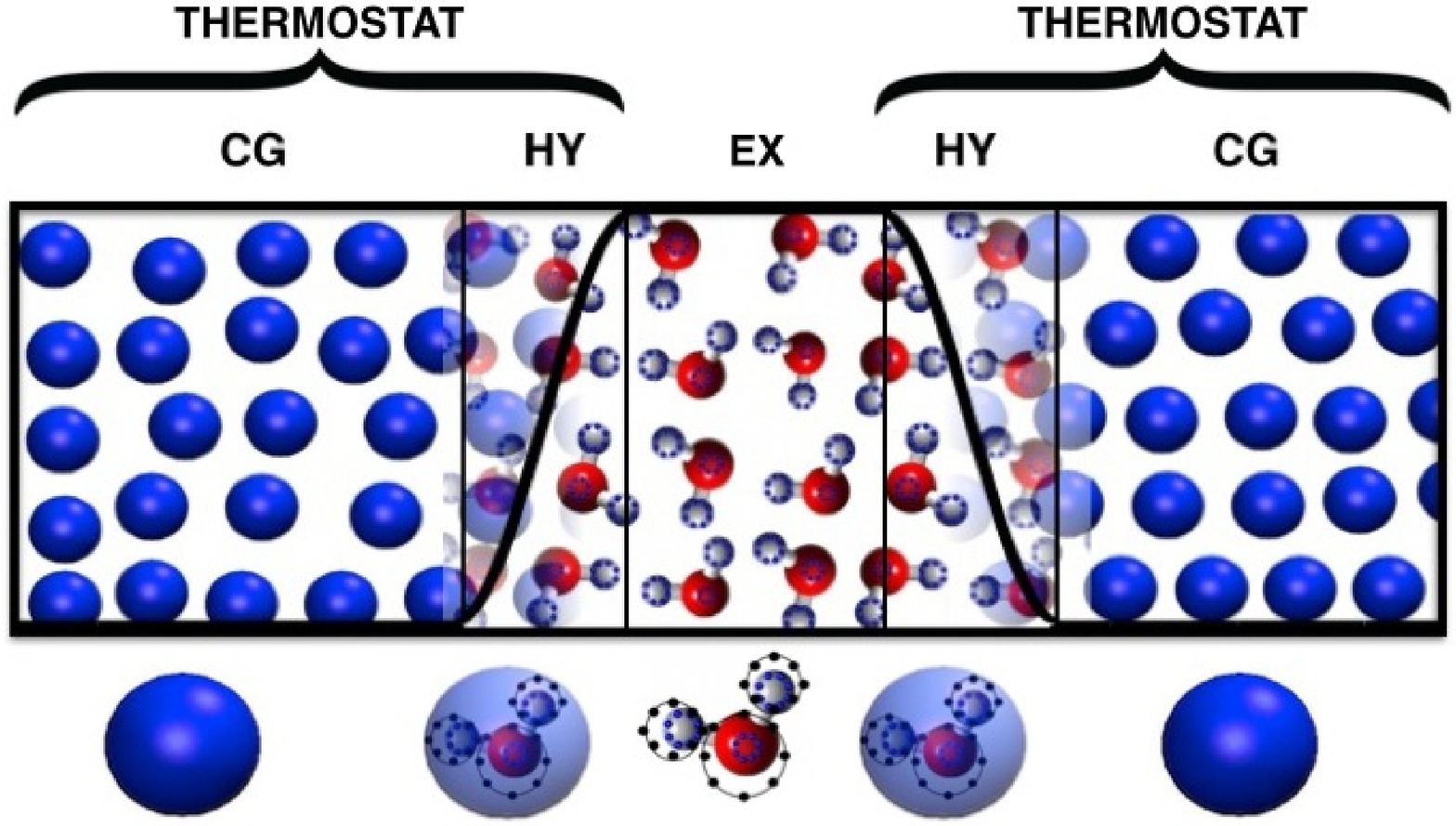}
  \caption{ Pictorial representation of the GC-AdResS scheme; CG indicates the coarse-grained region, HY the hybrid region where path-integral and coarse-grained forces are interpolated via a space-dependent, slowly varying, function $w(x)$ and EX (or PI) is the path-integral region (that is the region of interest). Top, the standard set up with the thermostat that acts globally on the whole system, used in the calculation of static properties. Bottom, the ``local'' thermostat technique employed in this work in the calculation of dynamical properties. }
  \label{fig1}
\end{figure}

In the original AdResS the coupling idea is rather simple, that is, in a region of interest (the atomistic or high resolution region) all the molecular degrees of freedom are treated via molecular dynamics while in a (larger) region of minor interest only coarse-grained degrees of freedom are treated. The passage of a molecule from one region to another should be performed smoothly with a hybrid dynamics in such a way that the atomistic and the coarse-grained regions are not perturbed in a significant way. In order to do so the space is divided into three regions, the atomistic (high resolution) region, the coarse-grained region and an interfacial region where the atomistic degrees of freedom are transformed in coarse-grained and vice versa, we call this region hybrid region or transition region (see Fig.\ref{fig1}). 
The coupling is made via a space dependent force interpolation:
\begin{equation}
F_{\alpha \beta} = w(X_{\alpha})w(X_{\alpha})F_{\alpha\beta}^{atom} + [1 - w(X_{\alpha})w(X_{\alpha})]F_{\alpha\beta}^{cm}
\label{force}
\end{equation}
where $\alpha$ and $\beta$ indicate two molecules, and $w(X_{\alpha})$ and $w(X_{\beta})$ indicate the interpolating (weighting) function depending on the coordinate of the center of mass of the molecules $X_{\alpha}$ and $X_{\beta}$:
\begin{equation*}
    w(x) = \begin{cases}
               1               & x < d_{AT} \\
               cos^{2}\left[\frac{\pi}{2(d_{\Delta})}(x-d_{AT})\right]   & d_{AT} < x < d_{AT}+d_{\Delta}\\
               0 & d_{AT} + d_{\Delta}< x
           \end{cases}
\end{equation*}
where, $d_{AT}$ and $d_{\Delta}$ are size of atomistic and hybrid regions respectively .
$F_{\alpha\beta}^{atom}$ is the force in the atomistic region, which is derived 
from atomistic interactions, $F_{\alpha\beta}^{cm}$ is the force in the coarse-grained region, which is derived from a coarse-grained potential. A thermostat takes care of thermally equilibrating the atomistic degrees of freedom reintroduced in the transition region. This simple set up turned out to be computationally robust; the calculation of structural and thermodynamics property in AdResS compared with the calculations done in a subregion of equivalent size in a full atomistic simulation shows a highly satisfactory agreement for several test systems \cite{pre1,jcppol,wat1,wat2,fullerene,jctcdeb}. The computational robustness encouraged the investigation of the conceptual justification of the method on the basis of first principles of thermodynamics and statistical mechanics \cite{prefra,jpa}. This analysis first led to the introduction of a thermodynamic force acting on the center of mass of the molecules in the hybrid region. The thermodynamic force is based on the principle of uniformizing, to the atomistic value, the chemical potential of each (space dependent) resolution \cite{jcpsimon} and then to the derivation of such a thermodynamic force from a more general thermodynamic principle, that is from the balance of grand potential for two interfaced open systems \cite{prlgio}:
\begin{equation}
\left[P_{atom}+\rho_{o}\int_\Delta F_{th}(x) dr \right]V = P_{CG}V
\end{equation}
where $P_{atom}$ and $P_{CG}$ are the pressure of the atomistic and coarse-grained region, $\rho_{o}$ is the target density of the reference full atomistic simulation, $V$ the volume of the simulation box. The explicit calculation of $F_{th}(x)$ is reported in the next section.
Based on such derivation a step forward was done and AdResS was reformulated in terms of Grand Canonical formalism (GC-AdResS) where mathematical rigorous conditions were derived in order to assure that in the atomistic region the system samples a Grand Canonical distribution. Such condition, at the first order, have been shown to be equivalent to the use of the thermodynamic force \cite{prx,jcpanim}. Moreover the coarse-grained model can be arbitrarily chosen, without any reference to the atomistic model. Recent results \cite{njpyn} have embedded the scheme into the Grand Ensemble model of Bergmann and Lebowitz \cite{bl1,bl2} and introduced a local thermostat acting only in the coarse-grained and hybrid region. Such a formalization allows one to define, with well founded physical arguments, the Hamiltonian of the atomistic (high resolution) region as the kinetic energy plus the interaction energy of the molecules in the atomistic region only; this implies that the interaction with molecules outside can be formally neglected. The definition of the Hamiltonian allows then to properly define the procedure for the calculation of equilibrium time correlation functions; moreover, for the case of PI approach, this set up will provide a rigorous definition of the Hamiltonian of quantization. As it will be specified later on, there exists also a clear numerical argument that supports the definition of an accurate Hamiltonian in the high resolution region.
\section*{PIMD Techniques}
The path integral formalism of Feynman applied to molecular simulation/dynamics of molecular systems is a well established approach and thus here we will not report its formal derivation but only those aspects which are technically relevant for this specific study. A formal derivation and discussion of basic aspect of this approach can be found, in Refs.\cite{tuck1,tuckbook2} for example. The essential point of interest (in this paper) is the transformation, via path integral formalism, of a classical Hamiltonian of N distinguishable particles with phase space coordinate (${\bf p},{\bf r}$), mass $m_{j}$ (for the j-th particle) and interaction potential in space $V({\bf r}_{1},....{\bf r}_{N})$:
\begin{equation}
H=\sum_{j=1}^{N}\frac{{\bf p}^{2}_{j}}{2m_{j}}+V({\bf r}_{1},....{\bf r}_{N})
\label{ham1}
\end{equation}
into a quantized Hamiltonian which is formally equivalent to a Hamiltonian of classical polymer rings (atoms). The interatomic potential is distributed over the beads in such a way that each bead of a polymer ring interacts with the corresponding bead of another polymer ring. The intra-atomic interactions consists of harmonic potentials which couple each bead to the first neighbors in the chain. The fictitious dynamics of this polymeric liquid, with the spatial fluctuations/oscillations of the rings describing the quantum spatial delocalization of the atoms, allows for the calculation of quantum statistical properties of the atomic/molecular system. The quantized Hamiltonian takes the form:
\begin{equation}
H_{P}=\sum_{i=1}^{P}\left[\sum_{j=1}^{N}\frac{[{\bf p}^{(i)}]^{2}_{j}}{2m^{'}_{j}}+\sum_{j=1}^{N}\frac{1}{2}m_{j}\omega^{2}_{P}({\bf r}^{(i)}_{j}-{\bf r}^{(i+1)}_{j})^{2}+\frac{1}{P}V({\bf r}^{i}_{1},....{\bf r}^{i}_{N})\right]
\label{quantam}
\end{equation}
where $P$ is the number of beads of the polymer, $m^{'}_{j}=\frac{P m}{(2\pi \hbar)^{2}}$ and ${\bf p}^{i}$ are a fictitious mass and momentum respectively, $\omega_{P}= \frac{\sqrt{P}}{\beta\hbar}$ ($\beta=1/k_{B}T$) and $V({\bf r}^{i}_{1},....{\bf r}^{i}_{N})$ is the potential that acts between same bead index $i$ of two different particles. This set up allows to use molecular dynamics for the calculation of statistical properties. However the direct use of the Hamiltonian above has shown to lead to a highly non-ergodic dynamics and suffers from poor sampling problems in the extended phase space of polymer ring~\cite{tuckbook}, since there are a wide range of frequencies present. The highest frequency limits the time step to be used in the simulation which causes the low frequency modes to be poorly sampled. Thus either a very small time step or very long runs should be performed, starting from different initial conditions in order to overcome this problem. In order to circumvent the ergodicity problem, normal modes transformation is preferred~\cite{tuck1, berne1}. The basic idea is to decouple the harmonic spring term, so that only a single harmonic frequency remains in the dynamics, and the time step for the simulation can be adjusted accordingly. The whole procedure is based on a transformation of coordinates to normal mode coordinates and thus to the use of an effective Hamiltonian:
\begin{equation}
H_{P}=\sum_{i=1}^{P}\left[\sum_{j=1}^{N}\frac{p_{j}^{(i)^{2}}}{2m_{j}^{(i)^{'}}} + \sum_{j=1}^{N}\frac{1}{2}m_{j}^{(i)}\omega_{P}^{2}(x^{'})^{(i)^{2}}_{j} + \frac{1}{P}V\left({\bf r}_{1}^{(i)}(x^{'}_{1}),\ldots, {\bf r}_{N}^{(i)}(x^{'}_{N})\right)  \right]
\label{quant-ham-nm}
\end{equation}
where $\frac{1}{P}U\left(r_{1}^{(i)}(x^{'}_{1}),\ldots, r_{N}^{(i)}(x^{'}_{N})\right)$ is the potential that acts between same bead index $i$ of two different particles in terms of the normal mode coordinates $x^{'}_{1},.....x^{'}_{N}$.
\subsection{Choice of masses}
In the standard PIMD ~\cite{pimd1, pimd2}, the masses $m_{j}^{(i)^{'}}$ are chosen such that  all the internal modes have the same frequency and the sampling is efficient. Thus the choice of mass is:
\begin{align*}
m_{j}^{(i)^{'}}&=m_{j}\lambda_{j}^{i}, i=2,\ldots,P\\
m_{j}^{1^{'}}&=m_{j}
\end{align*}
where $m_{j}$ is the physical mass and $\lambda_{j}^{i}$ are the eigenvalues obtained by the normal mode transformation. This approach was used to calculate static properties and here we will use it, within GC-AdResS, for the same purpose.
We will refer to this approach as {\bf H1}. 
Craig and Manolopolous~\cite{man2} have developed ring polymer molecular dynamics (RPMD), which has been successfully shown to calculate time correlation functions; the choices of the masses in RPMD is as follows:
\begin{equation}
m_{j}^{(i)^{'}}=m_{j}
\end{equation}
In this work, we will employ this approach within GC-AdResS to calculate, in addition to static properties, time correlation functions; we will refer to it as {\bf H2} approach.
However, there exists an alternative formulation for RPMD~\cite{man1}. The classical Hamiltonian for RPMD is:
\begin{equation}  
H_{P}=\sum_{i=1}^{P}\left[\sum_{j=1}^{N}\frac{[{\bf p}^{(i)}]^{2}_{j}}{2m_{j}}+\sum_{j=1}^{N}\frac{m_{j}}{2\beta_{P}^{2} \hbar^{2}}({\bf r}^{(i)}_{j}-{\bf r}^{(i+1)}_{j})^{2}+V({\bf r}^{i}_{1},....{\bf r}^{i}_{N})\right]
\label{manolopolous}
\end{equation}
where $\beta_{P}=\beta/P$, which effectively means that the simulation is performed at $P$ times the original temperature. Moreover the 
harmonic bead-bead interaction and the potential energy are scaled by $P$ relative to Eq~\ref{quant-ham-nm}. In Ref.~\cite{marx}, equivalence
between different RPMD formalisms was shown. 
Due to the calculation of the thermodynamic force, for GC-AdResS simulations this becomes an interesting technical aspect to investigate (see next sections). We will refer to this approach as {\bf H3} and verify its numerical robustness in GC-AdResS by comparison with the results obtained from {\bf H1,H2}.  
\subsection{PIMD in GC-AdResS}
The original idea of merging PIMD and AdResS was based on a simple extension of the AdResS principle. The dynamics of polymer rings, from a technical point of view, is nothing else than the dynamics of classical degrees of freedom, thus the standard AdResS could be applied (technically) in the same way, with only the modification \cite{prlado,para1,para2}:
\begin{equation}
F_{\alpha \beta} = w(X_{\alpha})w(X_{\alpha})F_{\alpha\beta}^{PI} + [1 - w(X_{\alpha})w(X_{\alpha})]F_{\alpha\beta}^{cm}
\label{piforce}
\end{equation}
where $F_{\alpha\beta}^{PI}$ is the force between beads of the rings representing the atoms of molecule $\alpha$ and molecule $\beta$.
However one of the authors has shown before that in {\bf any} adaptive scheme, based on a spatial interpolation of atomistic and coarse grained interactions, it cannot exist a valid rigorous global Hamiltonian \cite{premio}. Thus from the conceptual point of view the coupling between the polymer rings and the coarse-grained molecules cannot be rigorously expressed in a Hamiltonian form. However, calculations have shown that PIMD-AdResS was able to reproduce very well results obtained with full PIMD simulations.
Since the Hamiltonian formalism is at the basis of the PIMD approach the procedure of Refs.\cite{prlado,para1,para2} was empirical and could be verified only {\it a posteriori}. The reason why the procedure was successful is that the coupling between the polymer rings and the coarse-grained molecules is negligible, in terms of energetic contribution, under the hypothesis that the path integral region and the coarse-grained region were large enough compared to the hybrid region. However we have also numerically verified that even when all the three regions are relatively small and comparable in size, results are still satisfactory.
The latest formalization of AdResS in GC-AdResS, reported in the previous section, justifies why from a conceptual point of view the  setup of PI-AdREsS is robust. In fact according to the model of Bergmann and Lebowitz \cite{bl1,bl2,njpyn}, for a simulation in a Grand Ensemble one does not need to have an explicit coupling between the path integral region and the reservoir. The necessary and sufficient condition is the knowledge of the molecules' distribution in the reservoir. It follows that the interaction of the molecules of the path integral region with the rest of the system, while technically convenient and numerically efficient, from the conceptual (formal) point of view instead does not play a crucial role. Such interaction plays only the technical role of a sort of ``capping potential'' which avoids that molecules entering the path integral region overlap in space. 
Moreover the action of the thermodynamic force and of the thermostat in the hybrid region makes the stochastic coupling dominant (compared to the explicit hybrid interactions), which is the essence of any Grand Ensemble scheme. It follows that in GC-AdResS the Hamiltonian to be considered for the path integral formalism is the Hamiltonian of the path integral region only, without any external additional term; i.e. the path integral region with its quantized Hamiltonian is embedded in a large reservoir with the proper Grand Canonical behaviour.
It must be clarified that while the Bergmann-Lebowitz model provides an elegant and solid formal structure to the PI-AdResS, however it is not strictly required to justify the existence of an accurate Hamiltionian in the PI region and thus the implementation of PIMD in AdREsS.
 In fact in the appendix we provide a numerical proof that, for the systems treated in this paper, the interaction energy between the PI region and the rest of the system is at least one order of magnitude smaller than the interaction energy of the molecules in the PI region.
The accuracy and robustness of PI-AdResS (or PI-GC-AdREsS) will be shown with the simulation of liquid water in the next section. Finally it must be clarified that for the current implementation of PIMD in GC-AdResS (in the GROMACS package), it is difficult to estimate the computational gain since the code architecture is not yet optimized. At this stage we want only to show that the approach is satisfactory from a conceptual point of view. However, for very large systems with $P=32$, the computational gain is  around 1.7-2.0 compared to the full PIMD simulations. With further code modifications (e.g. removal of explicit degrees of freedom in the coarse-grained region, using multiple time steps) or with the implementation of PI-AdResS in a platform explicitly designed for PIMD simulation we estimated, for systems of the order of thousand molecules, a gain of at least a factor 4.0-5.0 compared to the full PIMD simulations.
\subsubsection{Calculation of the Thermodynamic Force in PIMD}
\begin{figure}[h!]
  \centering
  \includegraphics[width=0.75\textwidth]{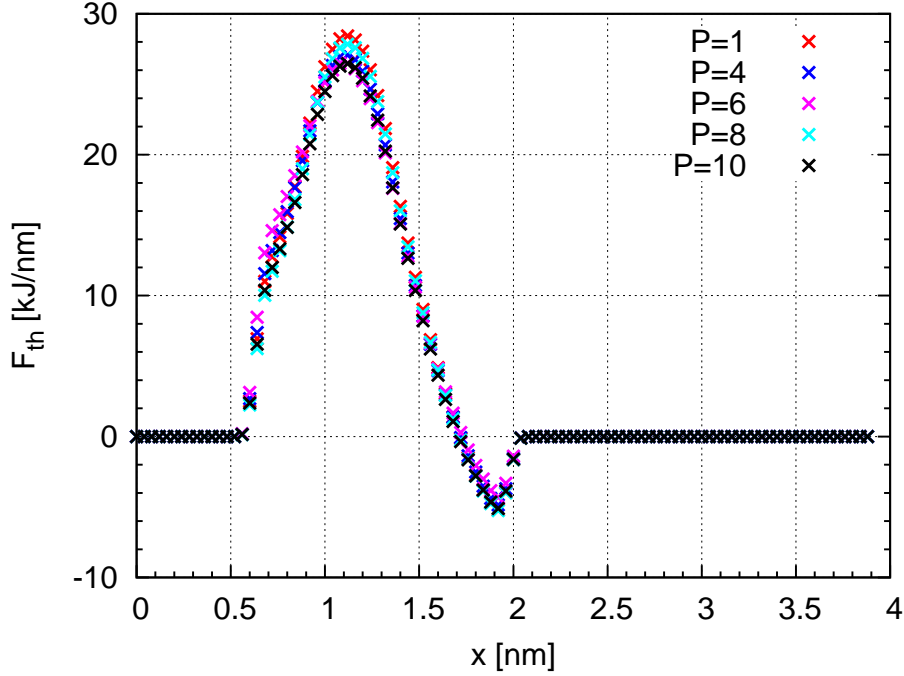}
  \caption{Thermodynamic force calculated in AdResS simulation using {\bf H1} ({\bf H2}) approach. The force is calculated for different number of polymer ring beads. It does not change as the number of beads is varied.}
  \label{thermomarx}
\end{figure}

For an atomistic system, the thermodynamic force, $F_{th}(x)$, can be expressed as:
\begin{equation}
F_{th}(x) = \frac{M}{\rho_{o}} \nabla P(x)
\end{equation} 
where $M$ is the mass of the molecule and $P(x)$ is the pressure which characterizes each different resolutions (for the initial configuration). $P(x)$ is approximated in terms of 
linear interpolation of molecular number density:
\begin{equation}
 P(x) = P_{atom} + \frac{M}{\rho_{o}\kappa}\left[\rho_{o} - \rho(x)\right]
\end{equation}
where $\rho(x)$ is the density generated if the simulation runs without any thermodynamic force. The thermodynamic 
force is then obtained by an iterative procedure:
\begin{equation}
F_{k+1}^{th}(x)=F_{k}^{th}(x) - \frac{M_{\alpha}}{[\rho_{o}]^2\kappa}\nabla\rho_{k}(x)
\end{equation}
After each iteration, a density profile $\rho(x)$ is obtained due to the application of the thermodynamic force. 
The process converges when the density profile obtained is equal to the target density. At this point the system is in thermodynamic equilibrium and the production run can start.
The calculation of thermodynamic force in PIMD-GC-AdResS is essentially based on the same principle of balancing grand potential for interfaced open systems:
\begin{equation}
\left[P_{quantum}+\rho_{o}\int_\Delta F_{th}(r) dr \right]V = P_{CG}V
\end{equation}
where $\rho_{o}$ is the target density of the reference full path-integral system.
As for the classical case, $P(x)$ can be written as:
\begin{equation}
 P(x) = P_{quantum} + \frac{M_{a}}{\rho_{o}\kappa}\left[\rho_{o} - \rho(x)\right]
\end{equation}
While the above approach is highly efficient for classical simulations, for path integral simulations, 
it is cumbersome to run an PIMD-GC-AdResS simulations to calculate the thermodynamic force, before doing an actual 
production run, as the path-integral simulation is inherently very expensive. In order to make the scheme efficient we have devised a strategy to calculate the thermodynamic force which requires least 
computation. As discussed in the  previous section, we will show how the thermodynamic force is calculated for the different Hamiltonian approaches. 
In case of {\bf H1}, and {\bf H2} where the temperature of the system is just the normal temperature, we calculated the thermodynamic
force for path-integral systems with varying Trotter number $P=1, 4, 6, 8$ and $10$ ($P=1$ represents the classical 
limit). Since the thermodynamic force takes care of a thermodynamic equilibration and since the thermodynamic conditions (thermodynamic state point) of a classical and a quantum system are the same, we expect that the thermodynamic force calculated in the classical case ($P=1$) is sufficient to provide thermodynamic equilibrium in simulations where $P=32$ is used.
In fact we found that the thermodynamic force was same in all the cases. Fig~\ref{thermomarx} shows the thermodynamic force calculated for 
a water system, with different number of ring polymer beads in each case. Using this argument, we used this thermodynamic force
in the actual production run with $P=32$. We found that the density of water molecules in the 
full quantum subregion and the transition region is equal to the reference density of the water system at the same
thermodynamic conditions. Thus, in the {\bf H1} and {\bf H2} approach, if the quantum effects on the pressure of the system are not large, we can directly use the thermodynamic force calculated from the classical simulation.

\begin{figure}[ht]
\centering 
 \includegraphics[width=0.75\textwidth]{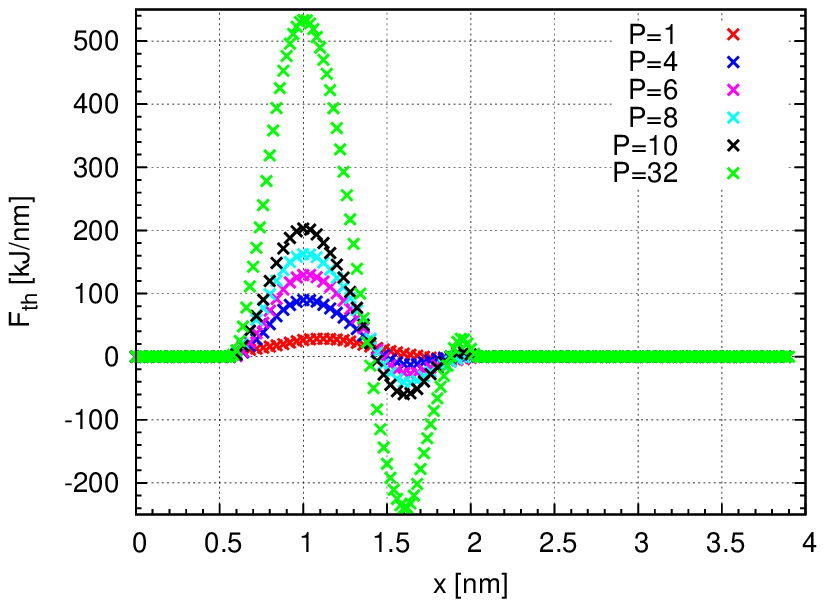}
  \caption{Thermodynamic force calculated in AdResS simulation using {\bf H3} approach. The force is calculated for different number of polymer ring beads. The 
  thermodynamic force for $P=32$ is then extrapolated by using space-dependent scaling factors calculated using thermodynamic force for $P=1, 4, 6, 8$ and $10$.}
\label{thermoman}
\end{figure}
In {\bf H3} approach, the situation is more complex, as the effective temperature of simulation
changes if the number of beads is changed, thus the (numerical) thermodynamic state point changes. In this case, there would be no other choice but to run a full PIMD-GC-AdResS simulation with $P=32$ and calculate the thermodynamic force. 
However, we avoided such an expensive calculation and instead calculated the thermodynamic force for system 
with different number of beads $P=1, 4, 6, 8$ and $10$ at temperatures $T=298\times P$ and extrapolated thermodynamic force for $P=32$, using 
space dependent factors calculated from thermodynamic force for smaller values of $P$.  Next we used this thermodynamic force 
for production run with $P=32$, and found that the density of water molecules in the full-PI subregion is same as the target
density while the density in transition region differs at worst by 3\%.
\subsubsection{Equilibrium Time Correlation Functions: Theoretical and computational aspects}
The technique of Ring Polymer molecular dynamics (RPMD) ({\bf H2}) focuses on the Kubo-transformed correlation functions \cite{kubo1,kubo2}. 
The Kubo-transformed correlation function of the operators $\hat{A}$ and $\hat{B}$ is defined by~\cite{man2}:
\begin{equation}
K_{AB}(t) = \frac{1}{\beta Z} \int_{0}^{\beta} d\lambda \left[e^{-(\beta - \lambda)\hat{\mathcal{H}}} \hat{A} e^{-\lambda \hat{\mathcal{H}}} e^{i\hat{\mathcal{H}}t / \hbar} \hat{B} e^{-i\hat{\mathcal{H}}t / \hbar}\right]
\end{equation}
where $Z$ is the canonical partition function:
\begin{equation}
Z=tr\left[e^{-\beta \hat{\mathcal{H}}}\right]
\end{equation}
The RPMD method approximates the Kubo-transformed correlation functions by using the classical ring-polymer trajectories 
generated by the dynamics produced by the Hamiltonian in Eq.~\ref{manolopolous}. The RPMD approximation is given by~\cite{manrev}:
\begin{equation}
\tilde{c}_{AB}(t) \approx \frac{1}{(2\pi\hbar)^{9PN}Z_{P}} \int \int d^{P}p_{0} d^{P}r_{0} e^{-\beta_{P} H_{P} (p_{0}, r_{0})} \frac{1}{N} \sum_{i=1}^{N} A^{i}_{P}(r_{0}) B^{i}_{P}(r_{t})
\end{equation} 
where $Z_{P}$ is the canonical partition function, and $r_{t}$ indicates the time evolution at time $t$ of the positions.
The functions $A_{P}(r_{o})$ and $B_{P}(r_{t})$ are calculated by taking the average over the beads of the ring polymer:
\begin{equation}
A_{P}(r) = \frac{1}{P} \sum_{j=1}^{P} A(r), B_{P}(r) = \frac{1}{P} \sum_{j=1}^{P} B(r)
\end{equation}
For the calculations in GC-AdResS the above equation needs to be written in the formalism of the Grand Canonical ensemble:
\begin{equation}
\begin{split}
\tilde{c}_{AB}(t)  & \approx \frac{1}{(2\pi\hbar)^{9PN}Z^{GC}_{P}} \sum_{N} \int \int d^{P}p_{0}(N) d^{P}r_{0}(N) e^{-\beta_{P} H_{P}(N) (p_{0}(N), r_{0}(N)) - \mu N} \\
                           & \times \frac{1}{N^{'}}  \sum_{i=1}^{N^{'}} A^{i}_{P}(r_{0}(N)) B^{i}_{P}(r_{t}(r_{0}(N)))
\end{split}
\end{equation}
where $\mu$ is the chemical potential and $N^{'}$ is the number of molecules at time `0', that 
remain correlated  at time `t' (that is molecules which remain in the path integral region for the whole time within the time window considered); $Z^{GC}_{P}=\sum_{N} e^{\beta \mu N}Z_{P}$ is the grand-canonical partition function and $H_{P}(N)$ is the Hamiltonian of the (open) path integral region with $N$, instantaneous number of molecules. It must be noticed that the {\it a priori} knowledge of $\mu$ is not required, actually in GC-AdResS $\mu$ is automatically calculated by the equilibration procedure of the thermodynamic force (see also \cite{jcpanim}).
From the technical point of view  we have used the same calculation procedure as that of  Ref.~\cite{njpyn}, where equilibrium time correlation functions were calculated in the 
open subsystems using classical molecular dynamics.  
Such a principle is based on the definition of reservoir in the Bergmann-Lebowitz model, which implies that when a molecule leaves the system and enters the reservoir, it looses its microscopic identity and thus the corresponding correlations; thus, if a molecule which is present at time $t_{0}$, disappears from the 
system at time $t$ (i.e. moves into the reservoir), then the contribution of this molecule, outside the time window $[0,t]$, to the correlation function shall not be considered. In our previous work we have shown that such a principle is physically consistent on the basis of results of molecular simulations.
Since all the beads in a ring-polymer are treated as dynamical variables~\cite{marx}, there are no thermostats used in RPMD simulations. Thus, the 
simulations are performed under NVE conditions, with either starting configurations generated from massively-thermostated 
PIMD simulations~\cite{muser}, or re-sampling of momenta from Maxwell-Boltzmann distribution after every few picoseconds~\cite{man5}. In order to keep the dynamics of the 
beads Newtonian in the path-integral subregion of GC-AdResS, we use a ``local-thermostat'' procedure~\cite{noneqhan, njpyn}, where the thermostat 
is applied only in the coarse-grained and hybrid region, while the explicit path-integral region is thermostat-free. This ensures, that the 
molecules which are present in the path-integral subregion are not subject to any perturbation due to the action of the thermostat.   
\section*{Results}
In this section we report results about the simulation of liquid water at room conditions. The quantum model for liquid water used in this work 
is q-SPC/FW~\cite{qspcfw}. It was shown that the thermodynamic and dynamical properties calculated using this water model agree quite well with the experiment data. The section is divided in two parts, the first where static results (molecular number density across the system, radial distribution functions, probability distribution of the molecules) are reported, and the second where several equilibrium time correlation functions are calculated. Few further points must be mentioned as clarification to this study. The total volume of the PIMD-GC-AdResS box is the same in all simulations, while three different sizes of the region at PI resolution are used and the dimension of the transition region is kept always the same. The smallest size of the PI region represents the limiting case of a statistically relevant number of molecules treated with PI resolution. The largest size instead represents the limiting case of a reservoir (hybrid plus coarse-grained region) which is relatively small and thus it may be expected to not fulfill the conceptual requirement of being statistically large enough. We will show that even in these two limiting cases the method is computationally and conceptually robust. A second point to take into account is that we compare the results of GC-AdResS for the PI region with the results obtained in a subsystem of a full PI simulations, such a subsystem is of the same size of the GC-AdResS simulation. The subsystem of a large full PI simulation box is a natural Grand Canonical ensemble, thus if our subsystem of AdResS reproduces the results of a full PI subsystem then we can be rather confident that the PI region in AdResS sample the Grand Canonical distribution sufficiently well.
From the physical point of view, it should be clarified that the functions calculated in a subsystem must be considered local in space and time if compared to calculation done over the whole simulation box of the full PI simulation. Once again, as the subsystem size increases the functions go to the value obtained in a full PI simulation when the full box is considered (physical consistency, see checks in Ref.\cite{njpyn}). Technical details of the simulation are reported in the Appendix.
\subsection{Static Properties}
\begin{figure}[!h]
  \centering
  \includegraphics[width=0.475\textwidth]{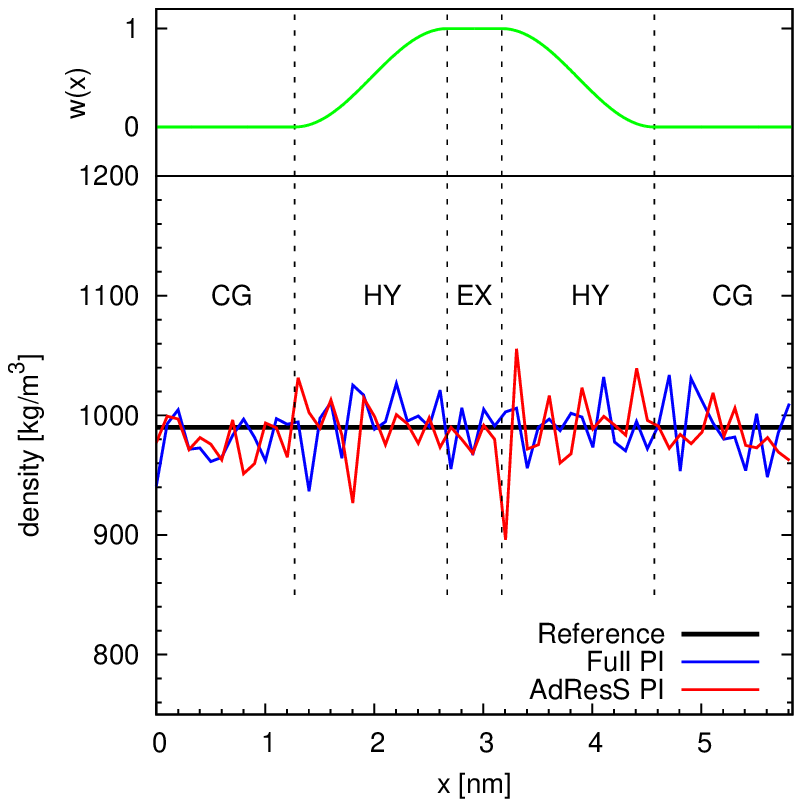}
  \includegraphics[width=0.475\textwidth]{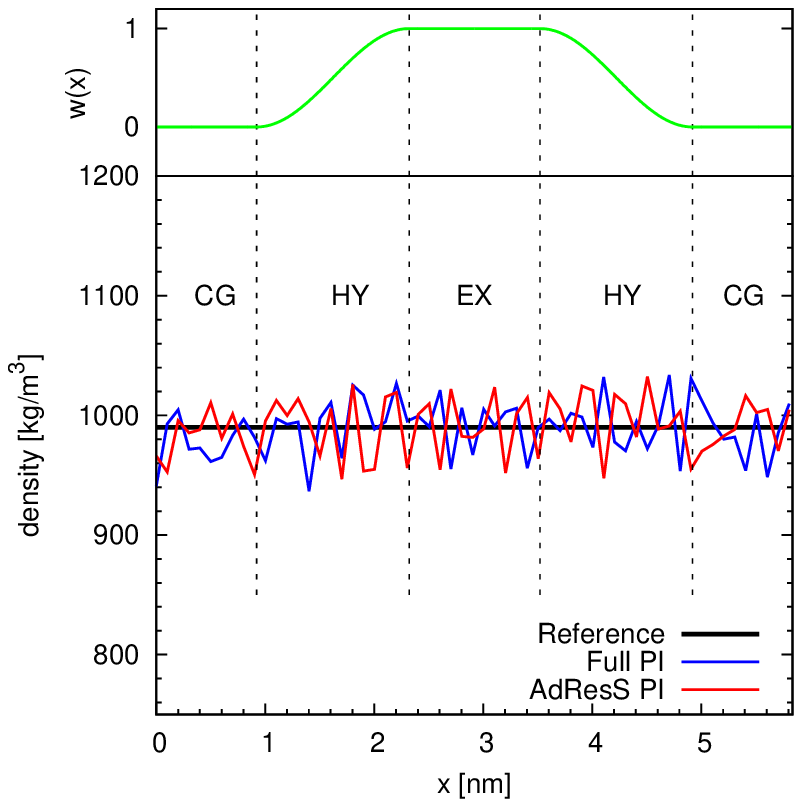}
  \includegraphics[width=0.475\textwidth]{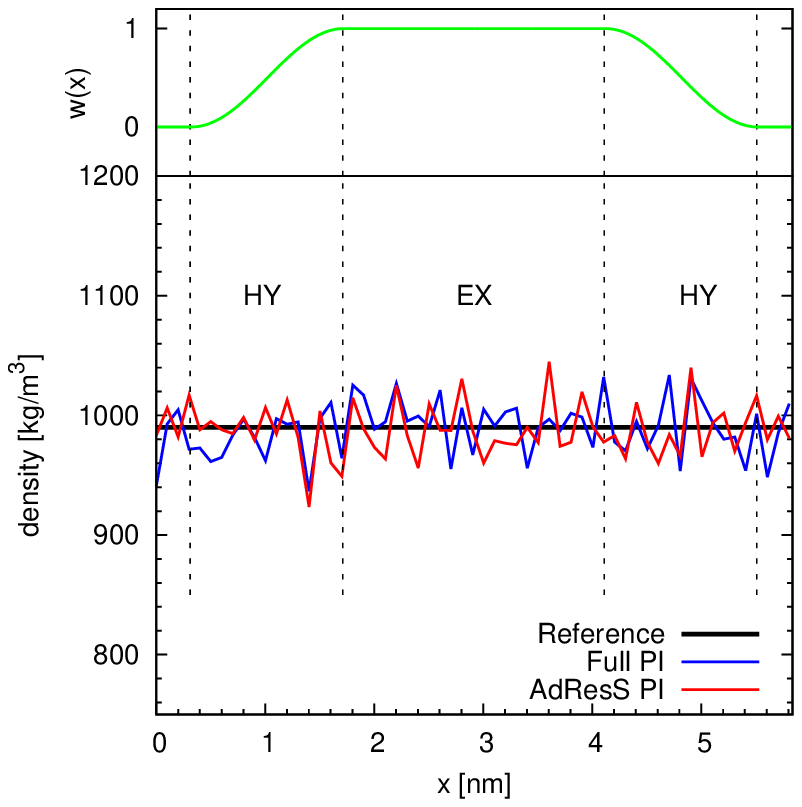}
  \caption{Molecular number density calculated with GC-AdResS for different size of quantum subregion. Results are compared with the density obtained in a full path integral simulation.}
  \label{density}
\end{figure}

\begin{figure}[!h]
  \centering        
  \includegraphics[width=0.475\textwidth]{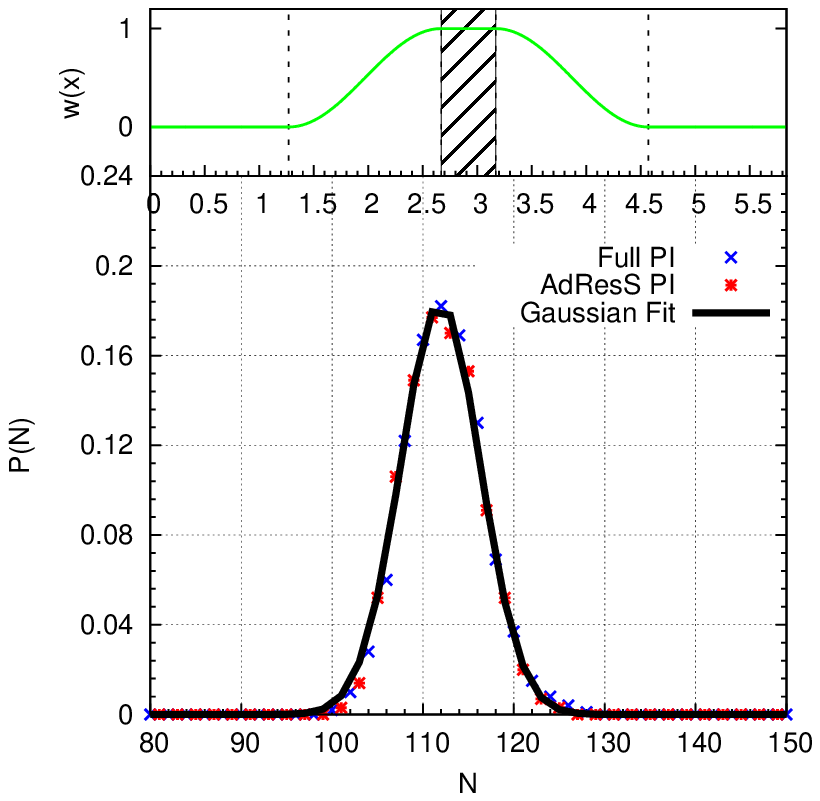}
  \includegraphics[width=0.475\textwidth]{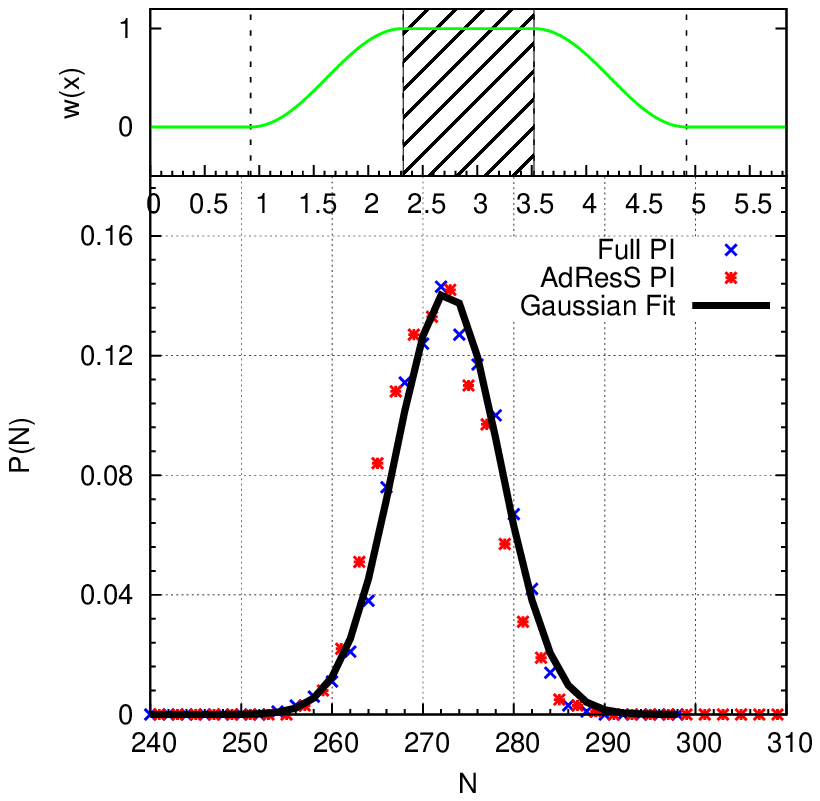}
  \includegraphics[width=0.475\textwidth]{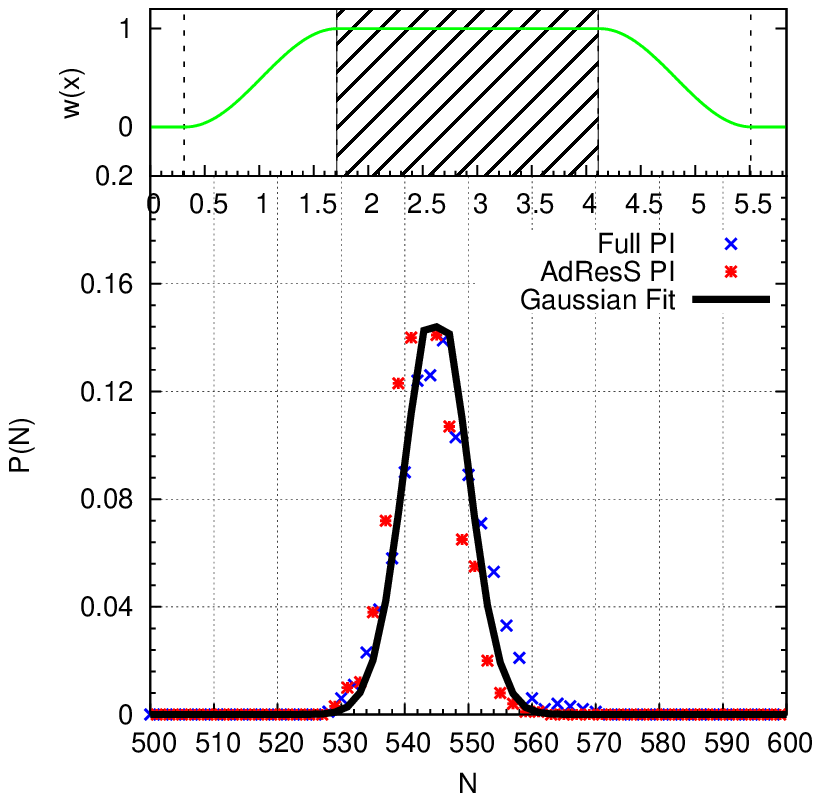}
  \caption{Particle number probability distribution of GC-AdResS compared with the equivalent full path integral subsystem, for different size of quantum subregion. The shape of both curves is a Gaussian (reference black continuous curve) in all the three different simulations. The top part of the figure indicates the extension of the PI region (compared to the rest of the system) where the function is calculated; this representation is equivalent in all subsequent figures.}
  \label{pn}
\end{figure}

\begin{figure}[!h]
  \centering
  \includegraphics[width=0.48\textwidth]{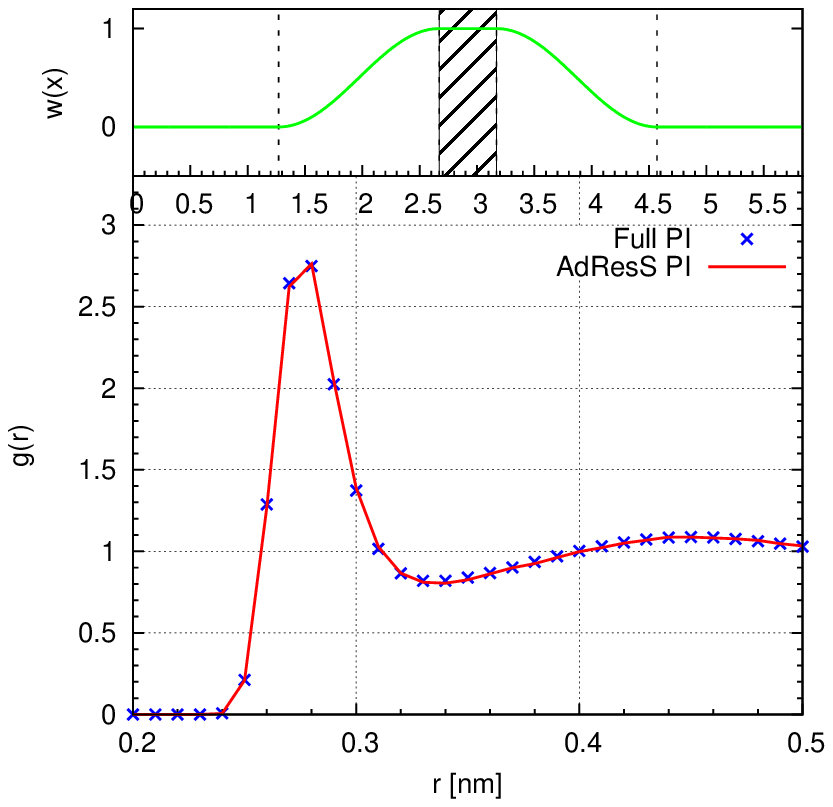}
  \includegraphics[width=0.48\textwidth]{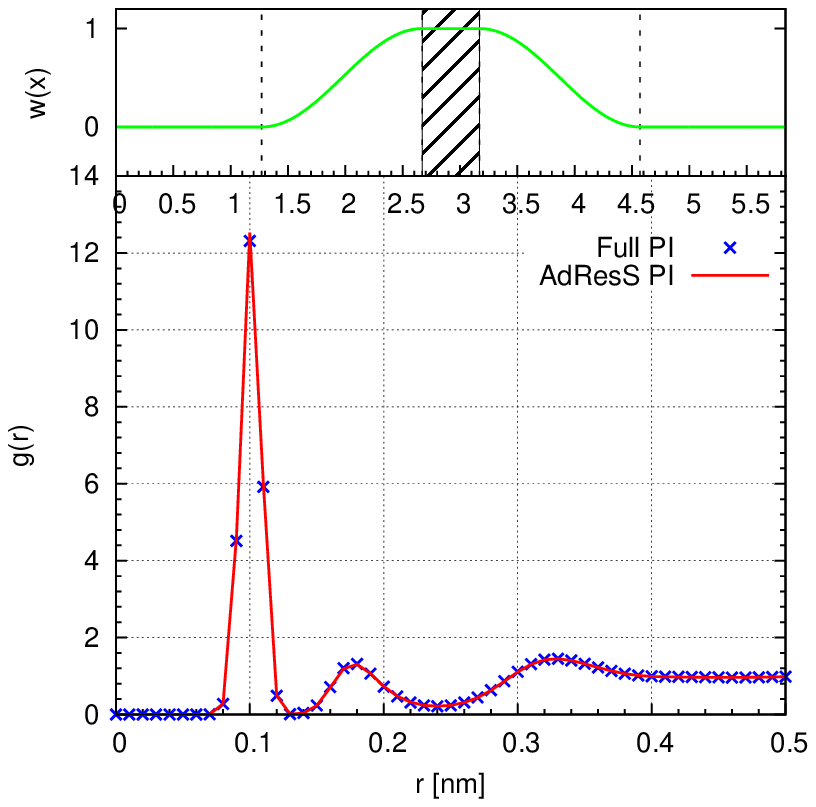}
  \includegraphics[width=0.48\textwidth]{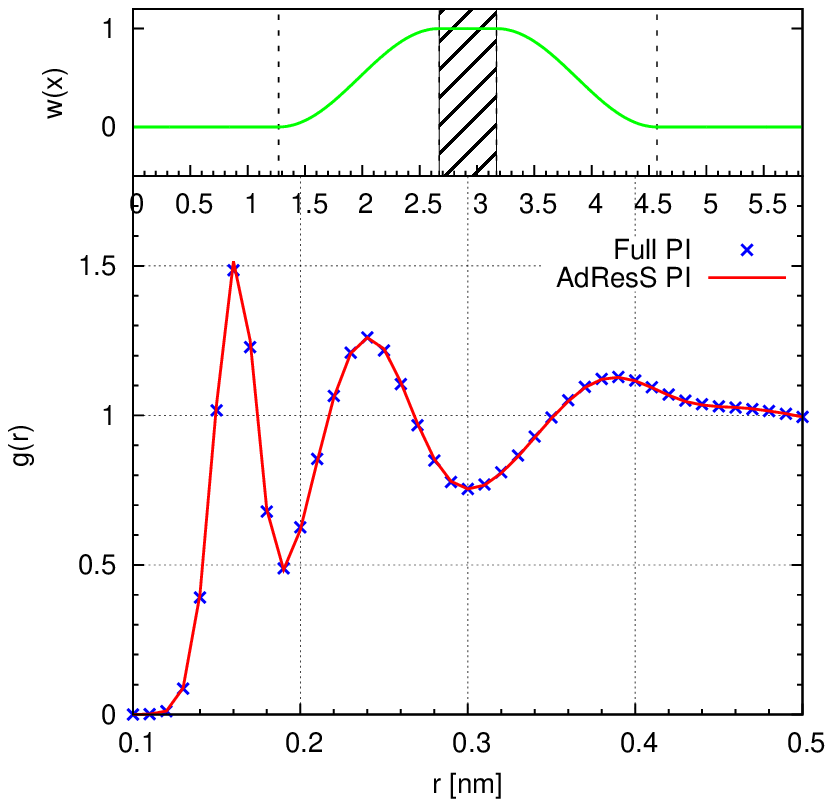}
  \caption{From left to right: (bead-bead) oxygen-oxygen, oxygen-hydrogen and hydrogen-hydrogen partial radial distribution functions calculated with path integral AdResS. Such functions are compared with the results obtained for an equivalent subsystem ($EX=0.5 nm$) in a full path integral simulation.}
  \label{gr1}
\end{figure}

We use the {\bf H1} and {\bf H2} PIMD approaches ({\bf H1}-GC-AdResS and {\bf H2}-GC-AdResS respectively for the GC-AdResS simulation), Fig.\ref{density} shows molecular number density. In all three cases the agreement is highly satisfactory, the largest deviation is found for the case with PI region of $0.5 nm$ and is below $5\%$. This is the basic test to show equilibration and thermodynamic consistency, moreover, following the mathematical formalization of Ref.\cite{prx} is the first order necessary condition in order to have the correct Grand Canonical distribution in the PI region.
A further confirmation of the fact that the method samples the phase space of a subsystem in a sufficiently correct way is represented by Fig~\ref{pn}.
The figure shows the particle number probability distribution in quantum subregion of AdResS and an equivalent subregion in full path integral simulation. It can be seen that also in this case the results are highly satisfactory and the shape of two curves is a Gaussian, as one should expect.
The g(r) is an important structural quantity that represents a two-body correlation function and thus a higher order than the molecular density of the ensemble many-body distribution; moreover it differs considerably when quantum models of water are used, in particular correlation functions involving hydrogen atoms~\cite{berne2}.   
We calculated the local bead-bead g(r)'s in the quantum subregion in GC-AdResS and compared them with the bead-bead g(r)'s in an equivalent subregion in the full path-integral simulation. Fig~\ref{gr1} to Fig~\ref{gr3} show that the results 
from GC-AdResS agree  with the results from full PI simulation in a highly satisfactory way. 

We have also verified, for the most relevant case ($EX=1.2$), that also the the {\bf H3} approach gives satisfactory results for the static properties when employed in GC-AdResS; results are reported in Fig~\ref{man}. Due to the more empirical calculation of the thermodynamic force in {\bf H3} results are not as accurate as those of {\bf H1} and {\bf H2}; the density in the hybrid region differs by around 3\%, which is anyway numerically negligible (however the difference must be reported).  However, the number probability distribution and bead-bead g(r)'s agree quite well in AdResS and full path-integral simulations. This leads to the conclusion that also results obtained with {\bf H3} are highly satisfactory. 

\begin{figure}[!h]
  \centering
  \includegraphics[width=0.48\textwidth]{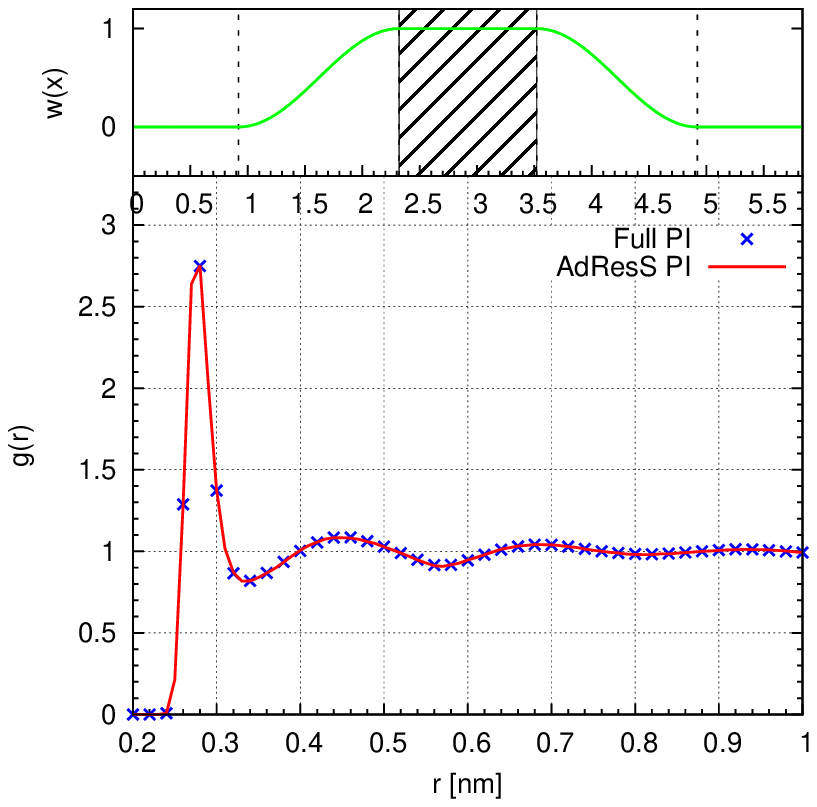}
  \includegraphics[width=0.48\textwidth]{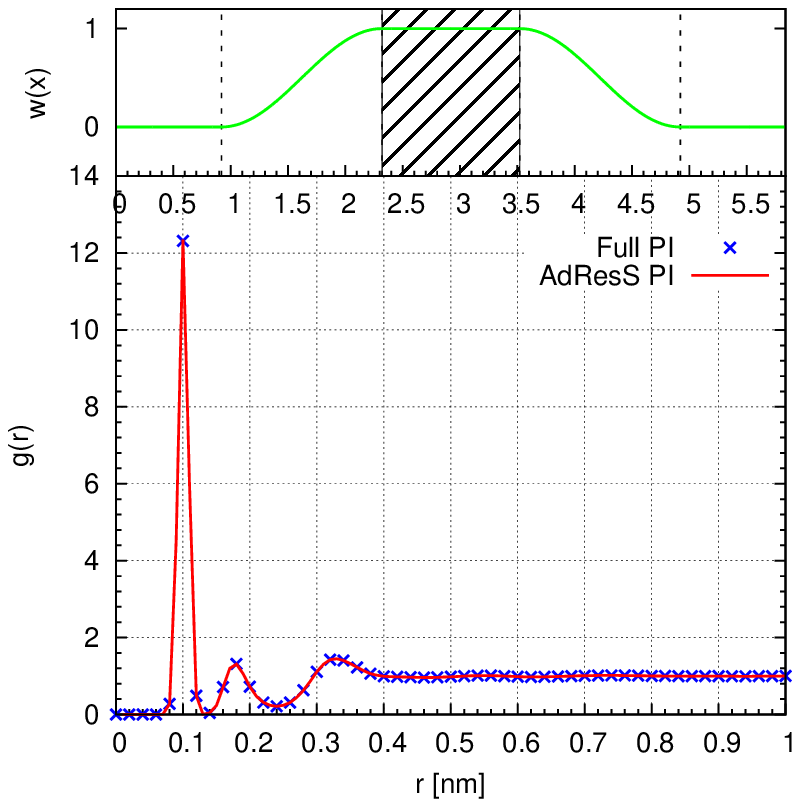}
  \includegraphics[width=0.48\textwidth]{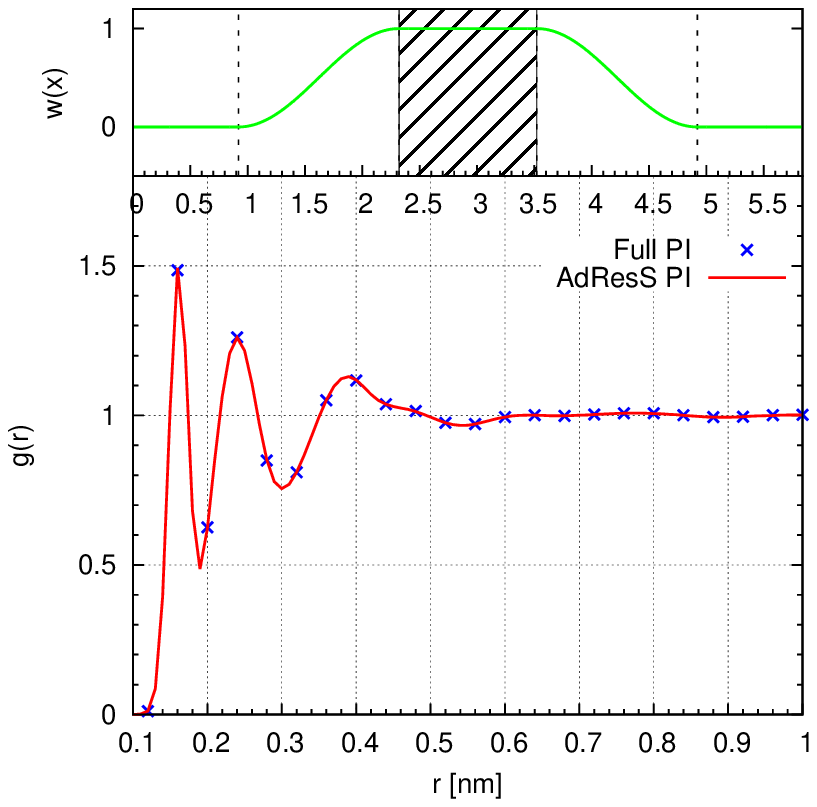}
  \caption{From left to right: (bead-bead) oxygen-oxygen, oxygen-hydrogen and hydrogen-hydrogen partial radial distribution functions calculated with path integral AdResS. Such functions are compared with the results obtained for an equivalent subsystem ($EX=1.2 nm$) in a full path integral simulation.}
  \label{gr2}
\end{figure}

\begin{figure}[!h]
  \centering
  \includegraphics[width=0.48\textwidth]{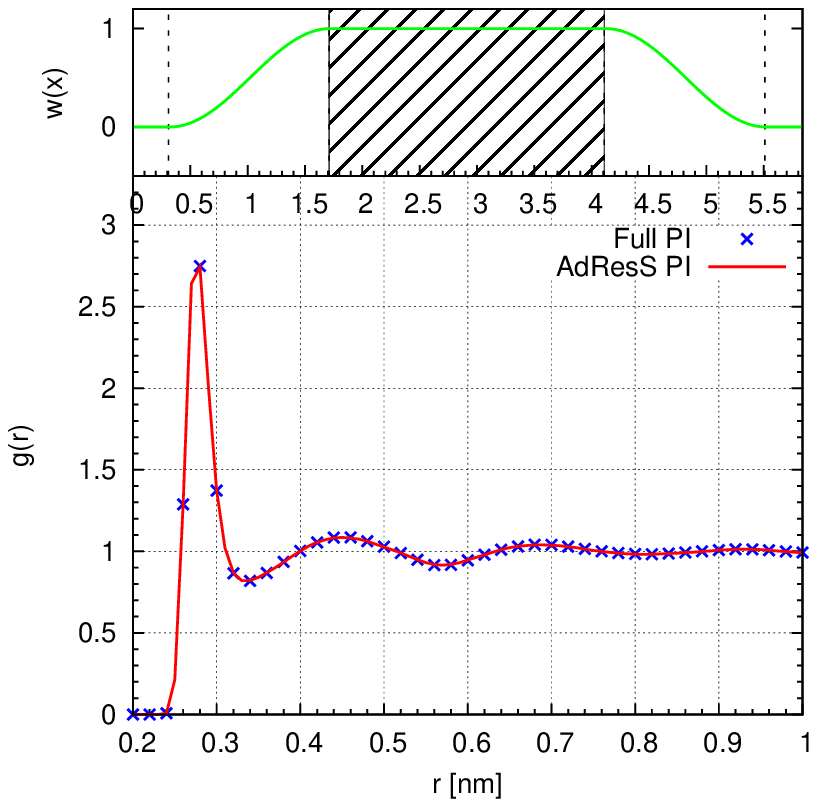}
  \includegraphics[width=0.48\textwidth]{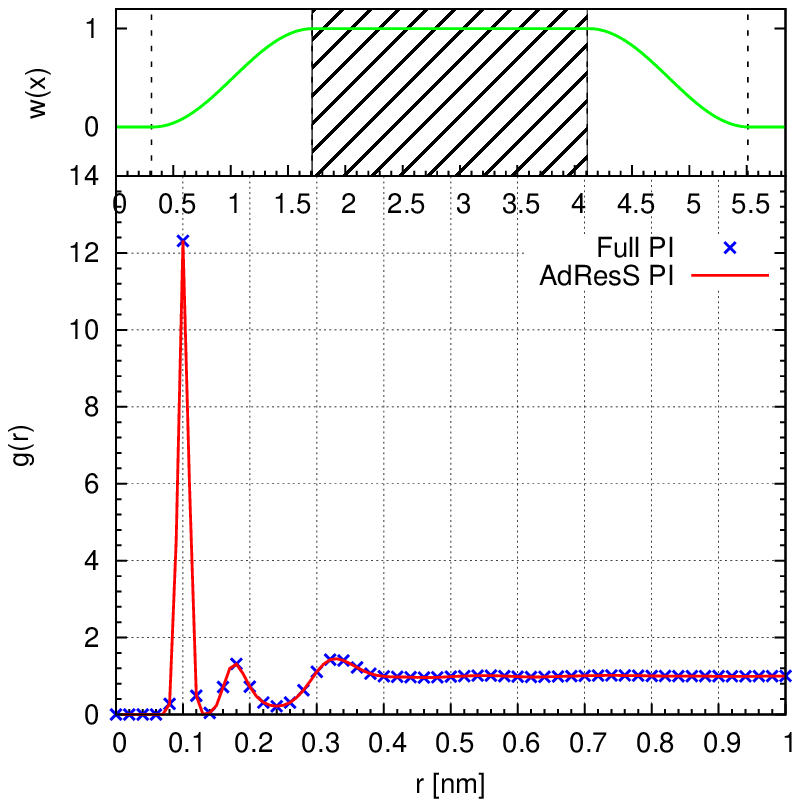}
  \includegraphics[width=0.48\textwidth]{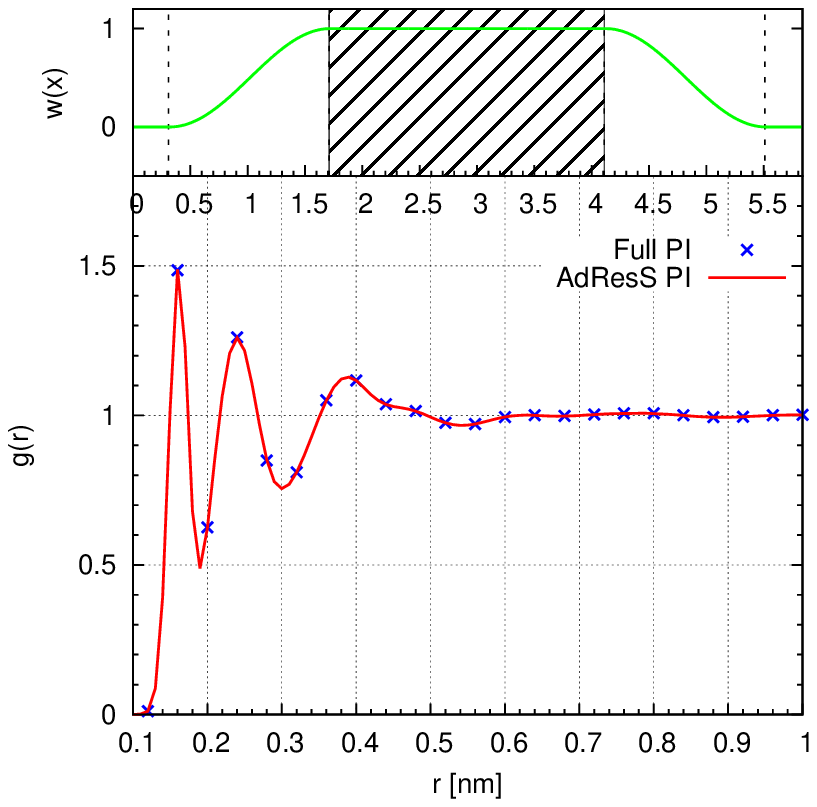}
  \caption{From left to right: (bead-bead) oxygen-oxygen, oxygen-hydrogen and hydrogen-hydrogen partial radial distribution functions calculated with path integral AdResS. Such functions are compared with the results obtained for an equivalent subsystem ($EX=2.4 nm$) in a full path integral simulation.}
  \label{gr3}
\end{figure}

\begin{figure}[!h]
  \centering
  \includegraphics[width=0.35\textwidth]{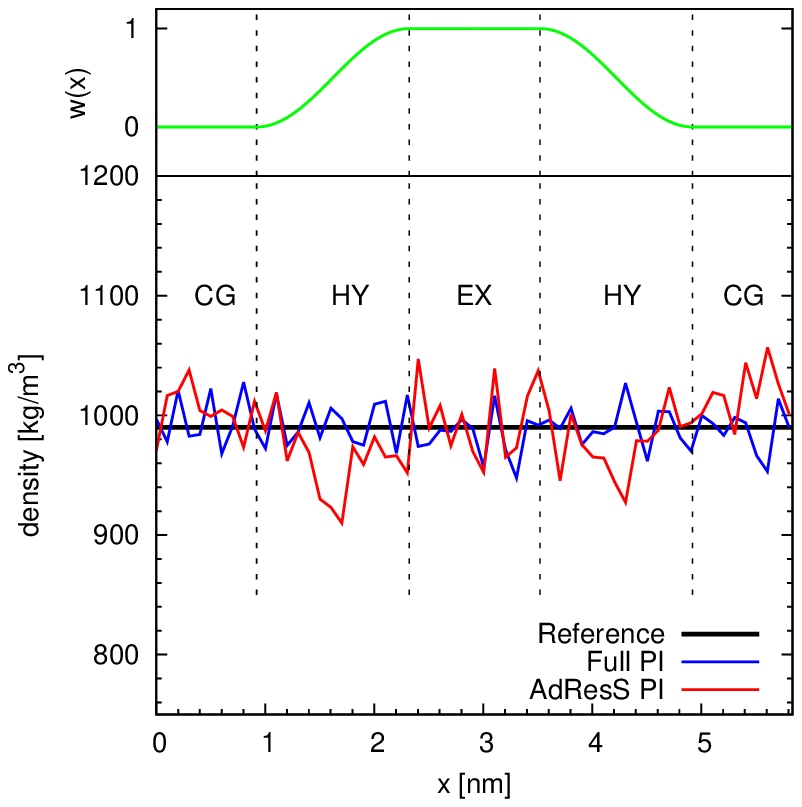}
  \includegraphics[width=0.35\textwidth]{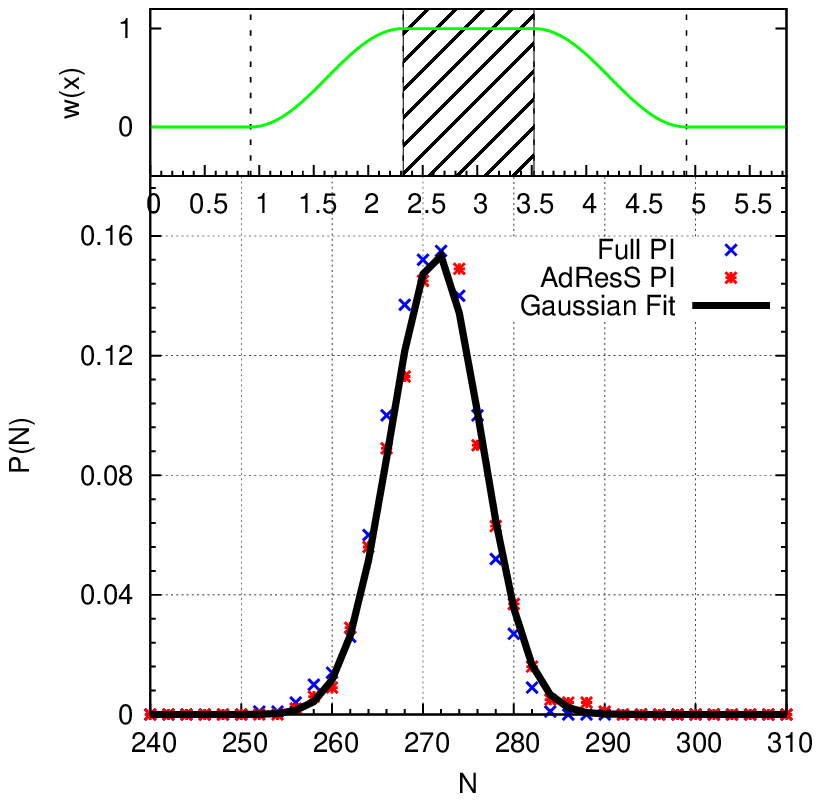}
  \includegraphics[width=0.35\textwidth]{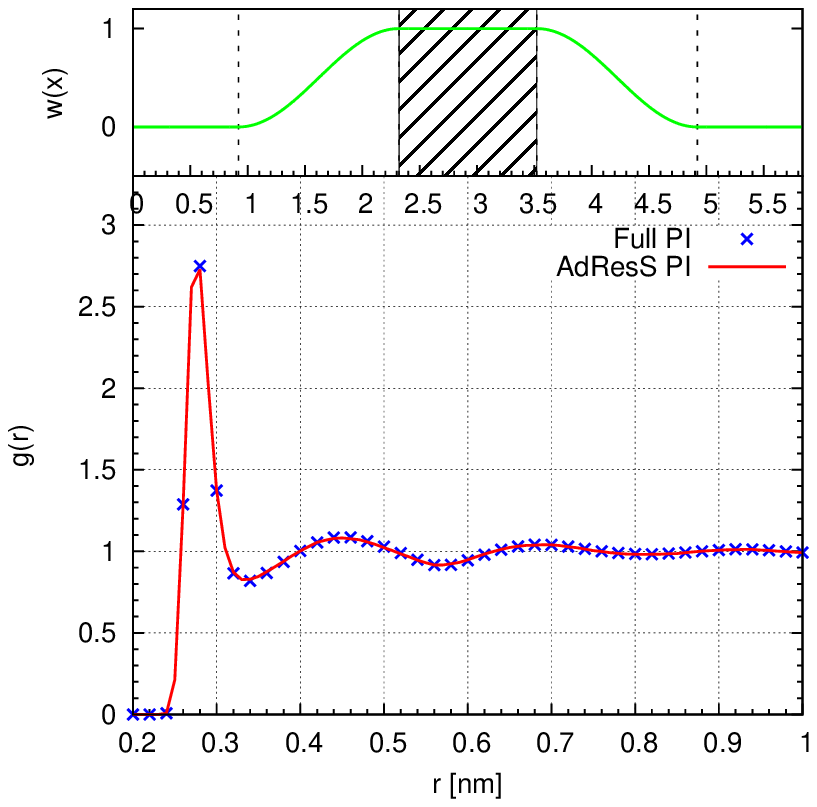}
  \includegraphics[width=0.35\textwidth]{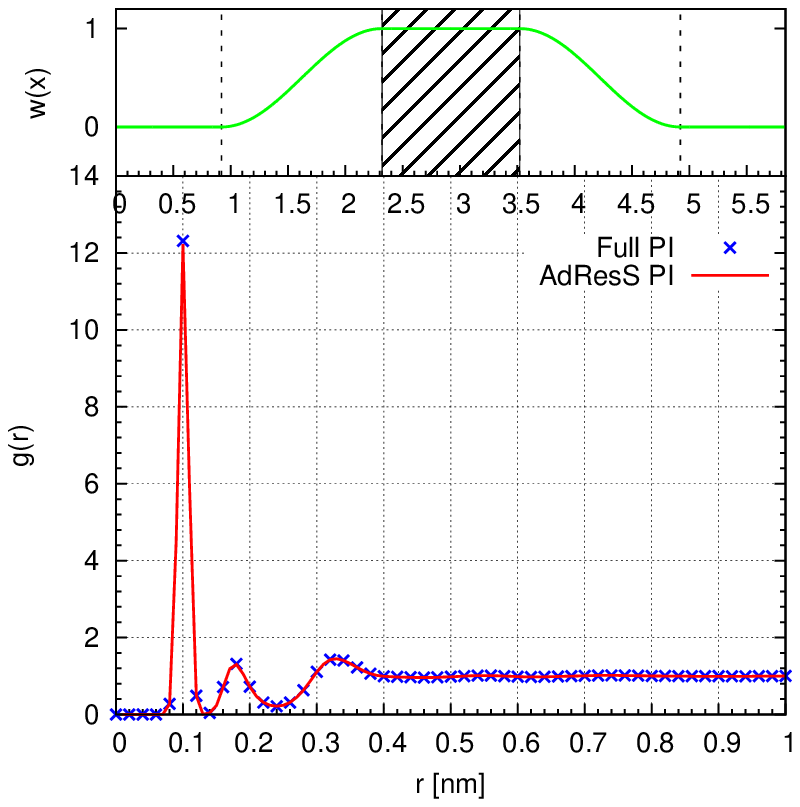}
  \includegraphics[width=0.35\textwidth]{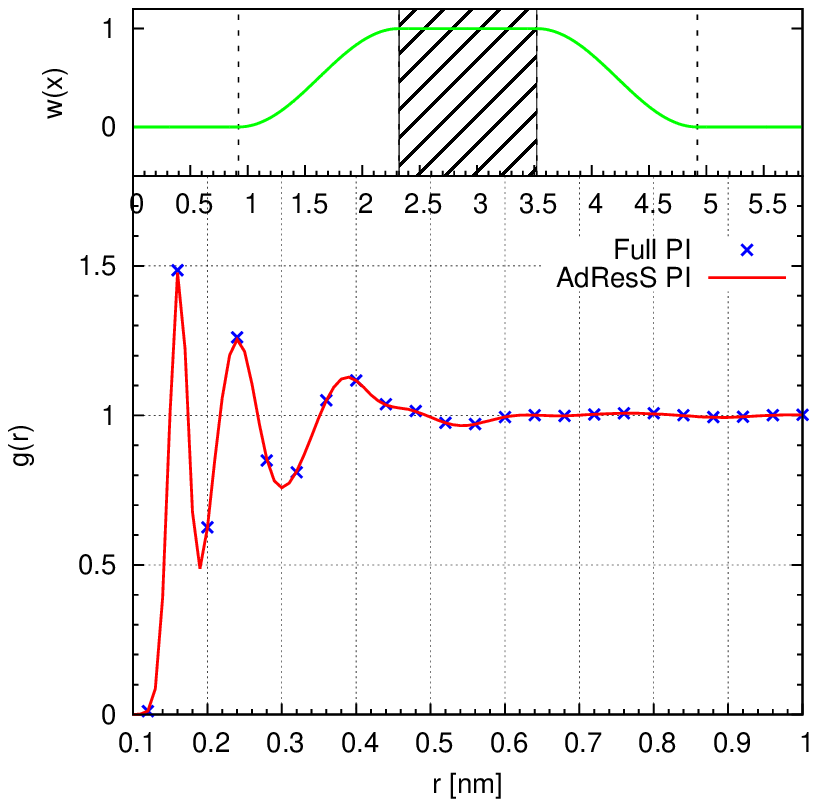}
  \caption{From left to right (top): Particle number probability distribution of GC-AdResS obtained using the {\bf H3} approach.\\
From left to right (bottom): (bead-bead) oxygen-oxygen, oxygen-hydrogen and hydrogen-hydrogen partial radial distribution functions calculated with path integral AdResS using the {\bf H3} approach. Such functions are compared with the results obtained for an equivalent subsystem ($EX=1.2 nm$) in a full path integral simulation.}
  \label{man}
\end{figure}
This section essentially show the ability of PI-GC-AdResS with all three PIMD techniques to sample basic (but highly relevant) static properties of a Grand Canonical ensemble. In order to prove that a more elaborated sampling is also satisfactorily made by the method we report in the next section the calculation of equilibrium time correlation functions.
\subsection{Dynamic properties}
 We report results for the velocity-velocity autocorrelation function, for the first and second order orientational (molecular dipole) correlation function \cite{orient1,orient2} and for the 
reactive flux correlation function for hydrogen bond dynamics \cite{chandler1,chandler2}.
This latter, in specific situations may strongly diverge from the classical case, and thus it may be a quantity of relevance; moreover the fact that PI-GC-AdResS reproduce the behaviour of a full PI simulation is of high technical relevance in perspective (e.g. study of solvation of molecules). 
The explicit formulas used for the functions calculated here are given in Ref.~\cite{man4}. All results shown in this section are highly satisfactory, either when {\bf H2} is used or {\bf H3} is used. Thus the PI-GC-AdResS can be certainly considered a robust computational method for the calculation of quantum-based static and dynamic properties of liquid water and as a consequence for simpler systems and for system where water play a major role (at least).
\subsubsection{Equilibrium Time Correlation Functions} 
\begin{figure}[!h]
  \centering
  \includegraphics[width=0.5\textwidth]{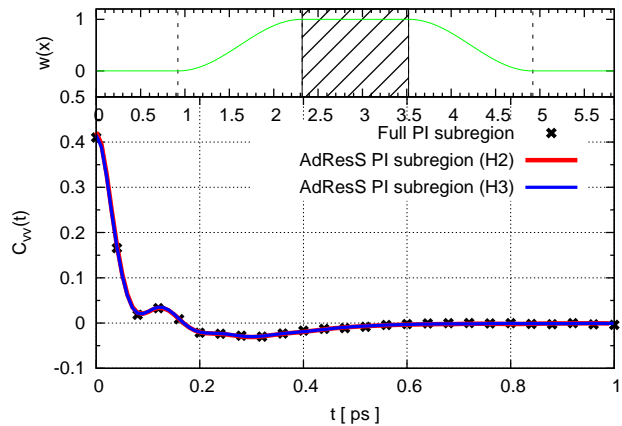}
  \caption{Kubo-transformed velocity auto correlation function for q-SPC/FW water model calculated in quantum subregion of GC-AdResS and
  an equivalent subregion in RPMD simulation.}
  \label{vac}
\end{figure}

\begin{figure}[!h]
  \centering
  \includegraphics[width=0.5\textwidth]{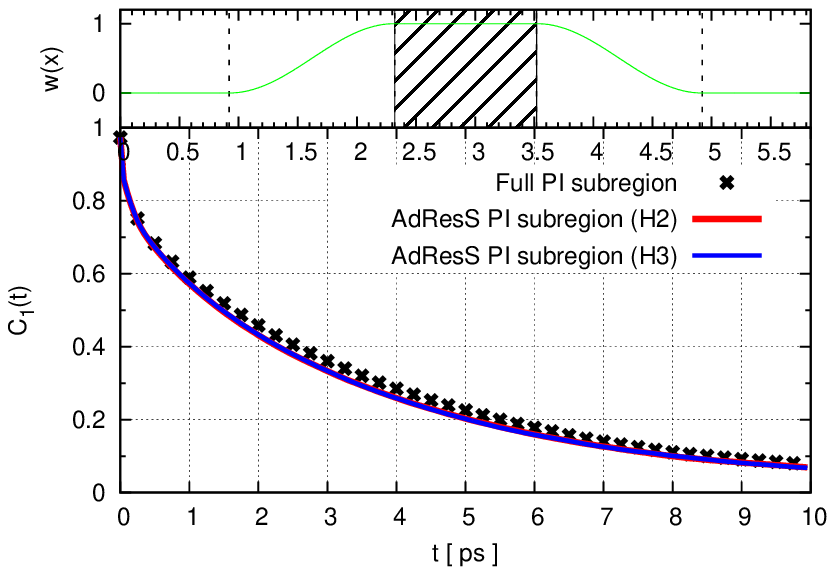}
  \caption{Kubo-transformed first order orientational correlation function for q-SPC/FW water model calculated in quantum subregion of GC-AdResS and
  an equivalent subregion in RPMD simulation. Dipole moment axis is chosen as the inertial axis of molecule.}
  \label{orient1x}
\end{figure}

\begin{figure}[!h]
  \centering
  \includegraphics[width=0.5\textwidth]{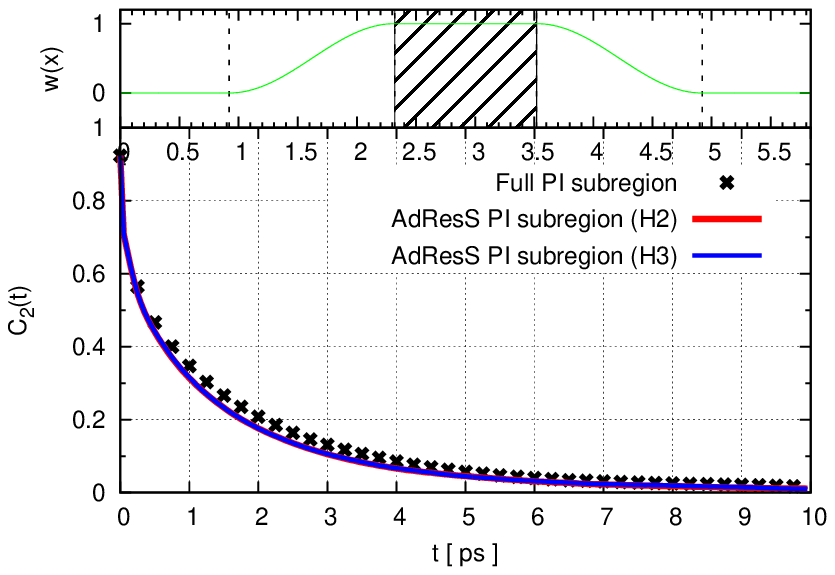}
  \caption{Kubo-transformed second order orientational correlation function  or q-SPC/FW water model calculated in quantum subregion of GC-AdResS and
  an equivalent subregion in RPMD simulation. Dipole moment axis is chosen as the inertial axis of molecule.}
  \label{orient2x}
\end{figure}

Figure~\ref{vac},~\ref{orient1x} and~\ref{orient2x} show the three correlation functions calculated in the quantum subregion in GC-AdResS and
an equivalent subregion in RPMD simulation, where the explicit region is 1.2 $nm$. All the 
correlation functions are calculated using {\bf H2} approach and {\bf H3}, which confirm the consistency of the two methods in GC-AdResS. As stated before, these are the local time correlation functions, calculated 
in the specific region of interest, and could differ from the global time correlation functions, calculated over the whole system. However, it 
was shown in Ref~\cite{njpyn}, that as the size of the explicit region increases, the local correlation functions converge to the global correlation functions.
\subsubsection{Dynamics of hydrogen bonding}
\begin{figure}[!h]
  \centering
  \includegraphics[width=0.5\textwidth]{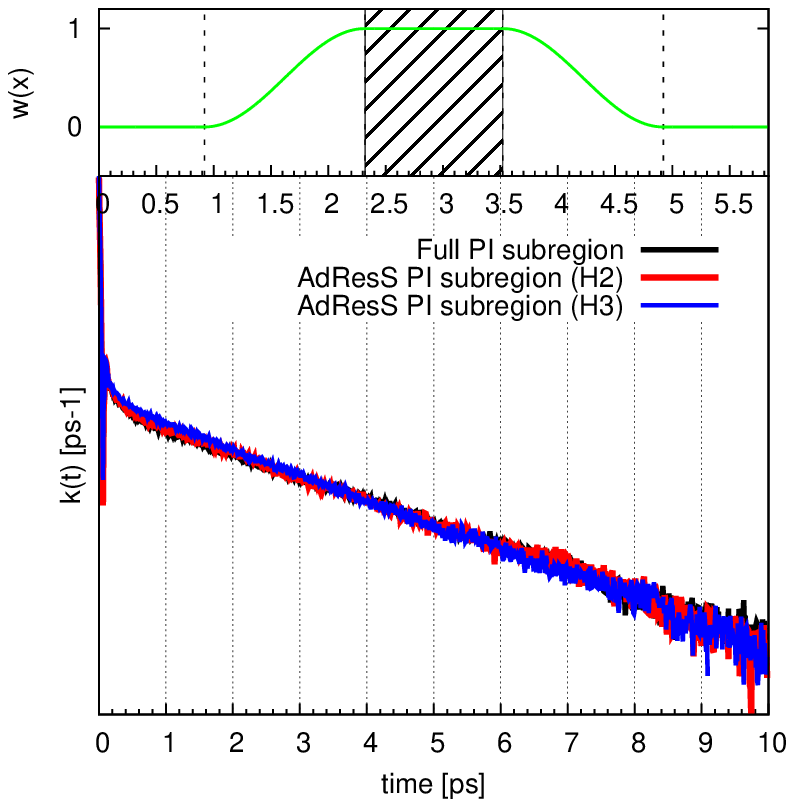}
  \caption{The rate function k(t) for q-SPC/FW water model calculated in quantum subregion of GC-AdResS and
  an equivalent subregion in RPMD simulation.}
  \label{hbond}
\end{figure}

In order to investigate the dynamics of hydrogen bond formation and breaking using RPMD simulations we calculate the hydrogen bond population fluctuations in time, which are characterized by the correlation function:
\begin{equation} 
c(t) = \langle h(0)h(t)\rangle/\langle h\rangle
\end{equation}
where $h(t)$ is the hydrogen bond population operator, which has a value 1, when a particular pair are bonded, and zero otherwise. 
One can then calculate the rate of relaxation as:
\begin{equation}
k(t) = -dc/dt 
\end{equation}
$k(t)$ is the average rate of change of hydrogen-bond population for those trajectories where the bond is broken at a time $t$ later. 
The two water molecules are treated as hydrogen bonded, if the distance between the center of two oxygen rings is less than 0.35 $nm$ and simultaneously 
the angle between the axis defined by the center of two oxygen ring polymers, and center of one of the oxygen-hydrogen ring is less than
$30$degrees. Fig~\ref{hbond} shows $k(t)$ calculated in the quantum subregion of AdResS and an equivalent subregion in 
RPMD simulation. 

\section*{Conclusion}
We have performed simulations of liquid water at room conditions using PIMD in three different technical approaches. Each of these approaches was embedded in GC-AdResS so that a PIMD for open systems in contact with a generic reservoir is realized. The results regarding static and dynamic quantities is highly satisfactory and qualified PI-GC-AdResS as a robust method for simulations of systems which currently are prohibitive for full PIMD simulations. For example the already mentioned solvation problem. One can define a high resolution region at PI resolution around the solute and surround the solvation region with a reservoir as that constructed in GC-AdResS. The static and dynamic properties of the hydrogen bonding network can be analyzed and, by comparing results with those of classical systems, one may conclude about the importance of quantum effects due to hydrogen spatial delocalization. This approach can introduce not only a technical innovation regarding the computational efficiency, but, by varying the size of the high resolution region, could be used as a tool of analysis to identify the essential degrees of freedom required by a certain physical process. In this perspective, here we have shown that PI-GC-AdResS is a  robust method for linking the microscopic to macroscopic scale in a truly multiscale fashion.  
\section*{Acknowledgments}
We thank Professor Thomas Markland for a critical reading of the manuscript and for useful suggestions. We thank Christoph Junghans, Han Wang and Adolfo Poma for many discussions during the formulation of the theory and the development of the code.We also thank Sebastian Fritsch for providing the path integral scripts.
This work was supported by the Deutsche Forschungsgemeinschaft (DFG) partially with the Heisenberg grant (grant code DE 1140/5-2) and partially with the grant CRC 1114 provided to L.D.S. The  DFG grant (grant code DE 1140/7-1) associated to the Heisenberg grant for A.G. is also acknowledged.
Calculations were performed using the computational resources of the North-German Supercomputing Alliance (HLRN), project {\bf bec00127}.

\section{Appendix: Technical details}
\subsection{Energetic contribution of the coupling term}
The $i$-th molecule (at position, $\vect r_i$) in the EX (PI) region is characterized by $w(\vect r_i)=1$. It follows that the force acting on the $i$-th molecule can be separated in two parts; (i) the force generated by the interaction of molecule $i$ with molecules of the EX region:
\begin{equation}
{\vect F}_{i,j}={\vect
  F}_{i,j}^{PI} , \forall j\in EX
\label{eqforceat}
\end{equation}
and (ii) the force generated by the interaction with molecules in the rest of the system:
\begin{equation}
{\vect F}_{i,j}=w(\vect r_j){\vect
  F}_{i,j}^{PI}+[1-w(\vect r_j)]{\vect F}^{CG}_{i,j} , \forall j\in HY+CG.
\label{eqforcehyb}
\end{equation}
From Eq.\ref{eqforceat} it follows:
\begin{equation}
{\vect F}_{i}=\sum_{j\neq i}{\vect F}_{i,j}^{PI}=\sum_{j\neq i}\nabla_{j}U^{ij}_{PI}
\label{grad1}
\end{equation}
where $\nabla_{i}$ is the gradient w.r.t. molecule $i$ and $U^{ij}_{PI}$ is a compact form to indicate the proper bead-bead interaction of atoms of molecule $i$ with those of molecule $j$.
Eq.\ref{eqforcehyb} represents instead the coupling force between molecules of $HY+CG$ region and molecule $i$, that is an external force. 
At this point we argue that the non-integrable part of the dynamics in the HY region is a numerically negligible boundary effect. In fact Eq.\ref{eqforcehyb} can be rewritten as:
\begin{equation}
{\vect F}_{i}=\sum_{j\in HY+CG}[w(\vect r_j){\vect
  F}_{i,j}^{PI}+[1-w(\vect r_j)]{\vect F}^{CG}_{i,j}]=\sum_{j\in HY+CG}[w(\vect r_j)\nabla_{i}U^{ij}_{PI}+[1-w(\vect r_j)]\nabla_{i}U_{CG}].
\label{grad2}
\end{equation}
It follows that the energy of the $i$-th molecule at a certain time time $t$ associated with the force of Eq.\ref{grad2} is given by: 
\begin{equation}
W^{i}_{PI-Res}(t)=\sum_{j\in HY+CG}[w(\vect r_j)U^{ij}_{PI}+[1-w(\vect r_j)]U^{ij}_{CG}], 
\label{waresat}
\end{equation}
where the $Res=HY+CG$. The  
 total energy of coupling at time $t$ is then defined as:
\begin{equation}
W_{PI-Res}(t)=\sum_{i\in PI}W^{i}_{PI-Res}(t)\,.
\label{toten}
\end{equation}
In order to understand whether or not the quantity of Eq.\ref{toten} is numerically negligible, one should compare it to the amount of energy, $W_{PI-PI}$, corresponding to the interaction between molecules of the PI region only: $W_{PI=PI}(t)=\sum_{i<j}U^{ij}_{PI}; i,j\in PI$.
If 
\begin{equation}
\frac{|W_{PI-PI}(t)|-|W_{PI-Res}(t)|}{|W_{PI-PI}(t)|}\approx 1; \forall t
\label{criter}
\end{equation}
 \begin{figure}
   \centering
   \includegraphics[width=0.5\textwidth]{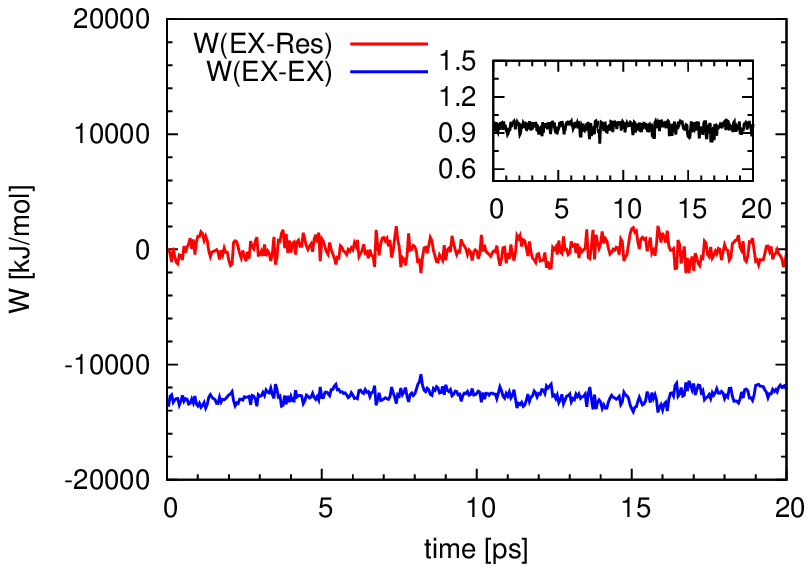}
   \includegraphics[width=0.5\textwidth]{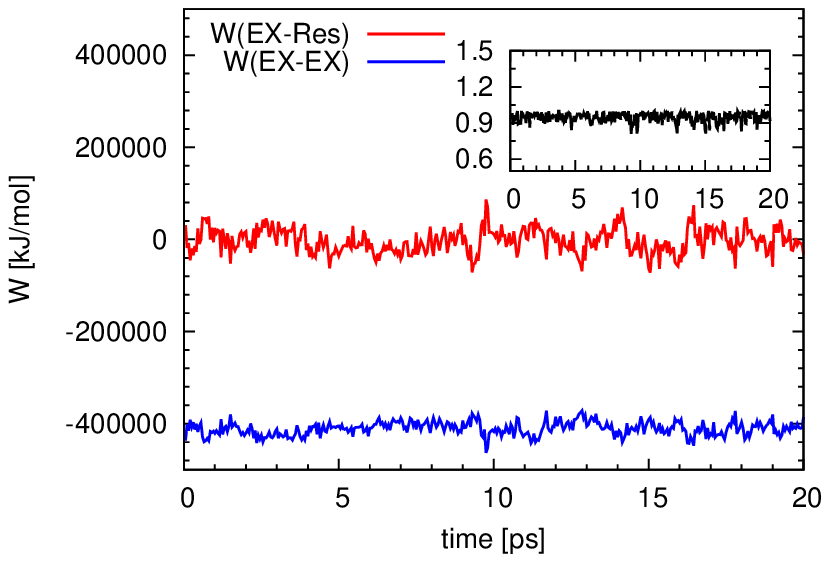}
   \caption{ Main figure: $W_{PI-PI}(t)$ compared to $W_{PI-Res}(t)$. Inset: The relative amount of the interaction between the PI region and the rest of the system along the trajectory: $\frac{|W_{PI-PI}(t)|-|W_{PI-Res}(t)|}{|W_{PI-PI}(t)|}$; the contribution is, at most, of $10\%$. Calculations are done within the {\bf H1} and {\bf H2} approach (top) and {\bf H3} (bottom).}
   \label{enpi1}
 \end{figure}
then it seems reasonable to approximate the total energy of the PI region by the Hamiltonian of the PI region, thus the Hamiltonian formalism is numerically justified in PI-AdResS.
Fig.\ref{enpi1} shows that the difference in energy is at least of one order of magnitude and that condition \ref{criter} holds in all simulations we have presented in this work.
Moreover it should be noticed that on purpose we have performed simulations where the technical conditions are not optimal (the size of each region of the system is much smaller than the size prescribed by the theory), thus Eq.\ref{criter} would certainly hold in simulations with standard technical conditions.
\subsection{simulation set up}
Static Properties: All path integral simulations are performed by home-modified GROMACS~\cite{gromacs}, and the thermodynamic force in 
GC-AdResS simulations is calculated using VOTCA~\cite{votca}.  The number of water molecules in system are 1320, and the 
box dimensions are $5.8\times2.6\times2.6$ $nm^3$, corresponding to a density $990$ $kg m^{-3}$. In AdResS simulations, 
the resolution of the molecules changes along x-axis, as depicted in Figure~\ref{fig1}. Three different AdResS simulations are 
performed, each with a different size of quantum subregion. The different sizes of the quantum subregion treated in 
this work are $0.5\times2.6\times2.6$ $nm^{3}$,  $1.2\times2.6\times2.6$ $nm^{3}$ and $1.2\times2.6\times2.6$ $nm^{3}$.
The transition region, which has dimensions $2.6 \times 2.6 \times 2.6$ $nm^{3}$ is fixed in all the three cases. The remaining system contains 
coarse-grained particles, which interact via generic WCA potential of the form:
\begin{equation}
U(r) = 4\epsilon\bigg[\bigg(\frac{\sigma}{r}\bigg)^{12} - \bigg(\frac{\sigma}{r}\bigg)^{6}\bigg]  + \epsilon,            r \leq 2^{1/6}\sigma
\end{equation}
The parameters $\sigma$ and $\epsilon$ in the current simulations are 0.30 nm and 0.65 kJ/mol respectively.
Thirty two ring polymer beads are used in all the simulations, which is sufficient to obtain converged results for both static and 
dynamical properties. Reaction field method is used to compute the electrostatic properties with dielectric 
constant for water = 80. The cut-off for both van der Waals and electrostatic interactions is 1.2 $nm$. All the static  
properties are computed from 250 ps long trajectories. The simulations using $H1$ and $H2$ formalisms are performed 
at 298 $K$, while the simulations using $H3$ formalism are performed at 9536 K. The time step used in all the simulations is 0.1 $fs$. 
In the calculation of thermodynamic force, a single iteration consists of a 200 ps long trajectory which is used to compute the density 
profile. A total of 20 such iterations is sufficient to obtain a flat density profile, and a converged thermodynamic force.

Dynamic Properties: The system details are kept same as in the previous section. A 200 ps long PIMD simulation is performed and along the trajectory, configurations are taken after every 8 ps to perform RPMD simulations. Thus a total of 25 trajectories each of length 25 ps are generated. 
For the first 5 ps, we keep the thermostat switched on, in order to adjust the velocities as masses are different 
in PIMD and RPMD methods. After this initial equilibration run, the thermostat is switched off, and the NVE 
trajectories generated are used to compute various time correlation functions.  We use the same strategy for AdResS simulations, 
where a 200 ps long fully thermostated GC-AdResS PIMD simulation is performed, and 25 initial configurations are taken along 
this trajectory to perform GC-AdResS RPMD simulations. For the first 5 ps, the thermostat acts in the explicit as well as the 
hybrid and coarse-grained regions. After the short equilibration run, the thermostat is switched off in the explicit region, while 
the hybrid and coarse-grained region are kept under the action of the thermostat. The dynamic properties are calculated in the 
explicit region in the last 20 ps, i.e. excluding the equilibration run. The velocity auto-correlation
function is calculated for 1 ps, while the orientational correlation functions and reactive flux correlation functions for hydrogen bond dynamics are calculated for 10 ps in one single trajectory, and then averaged over all the trajectories. 

\subsection{thermostat issue}
It is well known that massive thermosetting is needed in the path integral 
simulations, as the forces arising due to the high frequencies in the polymer
ring and the forces due to the potential $U(x)$ are weakly coupled. Tuckerman 
et al.~\cite{tuck1} coupled each normal mode variable to separate Nose-Hoover chains, 
thereby ensuring proper ergodic sampling of the phase space. Manolopolous et. al~\cite{man1} developed specific Langevin equations for thermostat, 
that are tuned to sample all the internal modes of the ring polymer quite efficiently. 
However, in this work we chose the standard Langevin equations of thermostat 
with time scale $0.1$ $ps$, which is strong enough for sampling the phase space 
effectively, though it may not be be the most efficient choice. The reason is that 
in the initial stage of validating GC AdResS for path integral simulations, we need to show that the properties 
obtained in the full PI simulations are reproduced exactly in AdResS. Since we use the same 
thermostat in both the simulations, there should not be any discrepancy arising due to the thermostat. However, the comparison of static properties calculated in our reference PIMD simulation with those available in literature (referring to the approaches above) is highly satisfactory.

\end{document}